# Synthetic Data in MR Spectroscopy: Current Practices, Applications, and Considerations

John T. LaMaster[1,2] 0000-0002-2149-771X†+ and Aaron T. Gudmundson[3] 0000-0001-5104-0959†+, Alireza Abaei[4] 0000-0003-2287-6495, Seyma Alcicek[5,6,7] 0000-0002-9447-4906, Arturo Alvarado[8] 0009-0006-1530-6389, Ovidiu Andronesi[9] 0000-0002-7412-0641, Tiffany Bell[10,11,12] 0000-0002-9591-706X, Wolfgang Bogner[13,14,15] 0000-0002-0130-3463, Hanna Bugler[10,11,12,16] 0009-0003-1331-3233, Alexander R. Craven[17,18] 0000-0003-2583-7571, Cristina Cudalbu[19,20] 0000-0003-4582-2465, Alma Davidson[21] 0009-0009-5040-1711, Christopher W. Davies-Jenkins[22] 0000-0002-6015-762X, Dinesh Deelchand[23,24] 0000-0003-4266-4780, Richard Edden[22] 0000-0002-0671-7374, Morteza Esmaeili[25] 0000-0003-0686-3571, Candace C. Fleischer[26,27] 0000-0003-3265-2276, Abdelrahman Gad[22] 0000-0002-4461-3517, Guglielmo Genovese[28] 0000-0001-9409-0120, Saumya Gurbani[26] 0000-0003-3524-3726, Ashley D. Harris[10,11,12] 0000-0003-4731-7075, Pierre-Gilles Henry[23,24] 0000-0002-8270-1956, Kay Chioma Igwe[21] 0000-0002-9909-9930, Ajin Joy[29] 0000-0003-3641-1218, Margarida Julià-Sapé[30,31,32] 0000-0002-3316-9027, Hyeonjin Kim[33,34] 0000-0003-2693-9983, Roland Kreis[35,36] 0000-0002-8618-6875, Fan Lam[37,38] 0000-0002-4124-0663, Karl Landheer[39] 0000-0001-5012-3007, Bernard Lanz[19,20] 0000-0001-5136-075X, Chu-Yu Lee[40] 0000-0001-5288-3309, Clémence Ligneul[41] 0000-0001-5673-3009, Julian P. Merkofer[42] 0000-0003-2924-5055, Jack J. Miller[43,44] 0000-0002-6258-1299, Jessie Mosso[19,45,46] 0000-0003-2924-5055, Stanislav Motyka[13,14,15] 0000-0002-6314-316X, Eloïse Mougel[47] 0009-0007-5926-2573, Paul G. Mullins[48] 0000-0002-1339-6361, Saipavitra Murali-Manohar[22] 0000-0002-4978-0736, Chloé Najac[49] 0000-0002-7804-2281, Shinichiro Nakajima[50,51] 0000-0002-2601-2195, Georg Oeltzschner[22] 0000-0003-3083-9811, İpek Özdemir[22] 0000-0001-6807-9390, Esin Ozturk-Isik[52] 0000-0002-8997-878X, Marco Palombo[53,54] 0000-0003-4892-7967, Ulrich Pilatus[55] 0000-0002-3207-7796, Justyna Platek[56,57,58] 0009-0008-0370-6304, Esau Poblador Rodriguez[59] 0000-0002-0524-2362, Xiaobo Qu[60,61,62] 0000-0002-8675-5820, Rudy Rizzo[63] 0000-0003-4572-5120, Christopher T. Rodgers[64] 0000-0003-1275-1197, Yeison Rodriguez[65] 0000-0002-8018-7148, Manoj K. Sammi[66] 0000-0002-3556-0468, Manoj Kumar Sarma[67] 0000-0002-5356-7276, Francesca Saviola[68] 0000-0001-7082-314X, Anouk Schrantee[69] 0000-0002-4035-4845, Amirmohammad Shamaei[70,71] 0000-0001-8342-3284, Dunja Simicic[22] 0000-0002-6600-2696, Brian J. Soher[72] 0000-0003-2750-7277, Nico Sollmann[73,74,75,76] 0000-0002-8120-2223, Yulu Song[22] 0000-0002-4416-7959, Jeffrey A. Stanley[77] 0000-0003-2355-480X, Bernhard Strasser[13] 0000-0001-9542-3855, Antonia Susnjar[78] 0000-0002-4973-5548, Kelley M. Swanberg[79,80] 0000-0002-1254-190X, M. Albert Thomas[29,81] 0000-0001-9037-2585, Ivan Tkáč[23] 0000-0001-5054-0150, Zhangren Tu[60,61,62] 0000-0002-0863-2454, Dennis M. J. van de Sande[82] 0000-0001-6112-1437, Emma Van Praagh[21] 0009-0008-7338-5673, Paul J. Weiser[9,13] 0009-0004-2503-5696, Mark Widmaier[19,20,83] 0009-0000-3353-0305, Martin Wilson[84] 0000-0002-2089-3956, Christopher J. Wu[21,85] 0009-0008-6829-1453, Lijing Xin[86] 0000-0002-5450-6109, Helge J. Zöllner[22] 0000-0002-7148-292X, MRS Synthetic Data Working Group, Antonia Kaiser[86] 0000-0001-7805-6766+*

+ Authors made equal contributions. † Shared first authorship. * Corresponding author.

---

**Affiliations:**

1. Munich Institute of Biomedical Engineering, Technical University of Munich, Munich, Germany
2. School of Computation, Information, and Technology, Technical University of Munich, Munich, Germany
3. The Malone Center for Engineering in Healthcare, Johns Hopkins University, Baltimore, MD, USA
4. ULMTeC Core Facility Small Animal Imaging Medical Faculty Ulm University, Ulm, Germany
5. Institute of Neuroradiology, University Hospital Frankfurt, Goethe University, Frankfurt am Main, Germany
6. University Cancer Center Frankfurt (UCT), Frankfurt am Main, Germany
7. Frankfurt Cancer Institute (FCI), Frankfurt am Main, Germany
8. Central University of Venezuela, Luis Razetti School of Medicine, Department of Pharmacology and Biochemistry, Caracas, Venezuela. CDD Las Mercedes
9. Athinoula A. Martinos Center for Biomedical Imaging, Department of Radiology, Massachusetts General Hospital, Harvard Medical School, Boston, MA, USA
10. Department of Radiology, University of Calgary, Calgary, AB, Canada
11. Hotchkiss Brain Institute, University of Calgary, Calgary, AB, Canada
12. Alberta Children's Hospital Research Institute, University of Calgary, Calgary, AB, Canada
13. High Field MR Center, Department of Biomedical Imaging and Image-Guided Therapy, Medical University of Vienna, Vienna, Austria
14. Christian Doppler Laboratory for MR Imaging Biomarkers (BIOMAK), Vienna, Austria
15. Comprehensive Center for AI in Medicine (CAIM), Medical University of Vienna, Vienna, Austria
16. Department of Biomedical Engineering, University of Calgary, Calgary, AB, Canada
17. Department of Biological and Medical Psychology, University of Bergen, Bergen, Norway
18. Department of Clinical Engineering, Haukeland University Hospital, Bergen, Norway
19. CIBM Center for Biomedical Imaging, Lausanne, Switzerland
20. École Polytechnique Fédérale de Lausanne (EPFL), Animal Imaging and Technology, Lausanne, Switzerland
21. Department of Biomedical Engineering, Columbia University, New York, NY, USA
22. Russell H. Morgan Department of Radiology and Radiological Science, Johns Hopkins School of Medicine, Baltimore, MD, USA
23. Center for Magnetic Resonance Research, University of Minnesota Medical School, Minneapolis, MN, USA
24. Department of Radiology, University of Minnesota Medical School, Minneapolis, MN, USA

25. Department of Electrical Engineering and Computer Science, Stavnger University, Stavanger, Norway
26. Department of Radiology and Imaging Sciences, Emory University School of Medicine, Atlanta, GA, USA
27. Department of Biomedical Engineering, Georgia Institute of Technology and Emory University, Atlanta, GA, USA
28. Department of Neuromedicine and Movement Science, Faculty of Medicine and Health Sciences, Norwegian University of Science and Technology, Trondheim, Norway
29. Department of Radiological Sciences, University of California, Los Angeles, Los Angeles, CA, USA
30. Departament de Bioquímica i Biologia Molecular, Universitat de Barcelona, Barcelona, Spain
31. Institut de Biotecnologia i Biomedicina (IBB), Universitat Autònoma de Barcelona, Barcelona, Spain
32. Centro de Investigación Biomédica en Red (CIBER), Barcelona, Spain
33. Department of Medical Sciences, Seoul National University, Seoul, Korea
34. Department of Radiology, Seoul National University Hospital, Seoul, Korea
35. MR Methodology, Department for Diagnostic and Interventional Neuroradiology, University of Bern, Bern, Switzerland
36. Translational Imaging Center (TIC), Swiss Institute for Translational and Entrepreneurial Medicine (sitem-insel), Bern, Switzerland
37. Department of Bioengineering, University of Illinois Urbana-Champaign, Urbana-Champaign, IL, USA
38. Beckman Institute of Advanced Science and Technology, University of Illinois Urbana-Champaign, Urbana-Champaign, IL, USA
39. Regeneron Genetics Center, Tarrytown, NY, USA
40. Department of Radiology, University of Iowa, Iowa City, IA, USA
41. Wellcome Centre for Integrative Neuroimaging, FMRIB, Nuffield Department of Clinical Neurosciences, University of Oxford, Oxford, UK
42. Department of Electrical Engineering, Eindhoven University of Technology, Eindhoven, The Netherlands
43. The MR Research Centre, Aarhus University, Aarhus, Denmark
44. Clarendon Laboratory, Department of Physics, University of Oxford, Oxford, UK
45. CIBM Pre-Clinical Imaging EPFL Metabolic Imaging Section, École Polytechnique Fédérale de Lausanne, Lausanne, Switzerland
46. Bernard and Irene Schwartz Center for Biomedical Imaging, Department of Radiology, New York University Grossman School of Medicine, New York, NY, USA
47. Université Paris-Saclay, Commissariat à l'Energie Atomique et aux Energies Alternatives (CEA), Centre National de la Recherche Scientifique (CNRS), Molecular Imaging Research Center (MIRCen), Laboratoire des Maladies Neurodégénératives, Fontenay-aux-Roses, France
48. School of Psychology and Sport Science, Bangor University, Bangor, Gwynedd, UK

49. C.J. Gorter MRI Center, Department of Radiology, Leiden University Medical Center, Leiden, The Netherlands
50. Department of Neuropsychiatry, School of Medicine, Keio University, Tokyo, Tokyo, Japan
51. Multimodal Imaging Group, Brain Health Imaging Centre, Centre for Addiction and Mental Health, Toronto, ON, Canada
52. Institute of Biomedical Engineering, Bogazici University, Istanbul, Türkiye
53. Cardiff University Brain Research Imaging Centre (CUBRIC), School of Psychology, Cardiff University, Cardiff, UK
54. School of Computer Science and Informatics, Cardiff University, Cardiff, UK
55. Institute of Neuroradiology, Goethe-University Frankfurt, Frankfurt am Main, Germany
56. Division of Medical Physics in Radiology, German Cancer Research Center (DKFZ), Heidelberg, Germany
57. Faculty of Physics and Astronomy, Heidelberg University, Heidelberg, Germany
58. International Max Planck Research School for Quantum Dynamics in Physics, Chemistry, and Biology, MPIK, Heidelberg, Germany
59. Competence Center for Preclinical Imaging and Biomedical Engineering, Faculty of Health, University of Applied Sciences, Wiener Neustadt, Austria
60. Department of Electronic Science, School of Electronic Science and Engineering, Xiamen University, Xiamen, China
61. Fujian Provincial Key Laboratory of Plasma and Magnetic Resonance, Xiamen University, Xiamen, China
62. Biomedical Intelligent Cloud Research and Development Center, National Institute for Data Science in Health and Medicine, Xiamen University, Xiamen, China
63. Department of Radiology, University of Michigan, Ann Arbor, MI, USA
64. Department of Clinical Neurosciences, University of Cambridge, Cambridge, UK
65. Advanced Imaging Research Center, Department of Biomedical Engineering, University of Texas Southwestern Medical Center, Dallas, TX, USA
66. Advanced Imaging Research Center, Oregon Health & Science University, Portland, OR, USA
67. Advanced Imaging Research Center, University of Texas Southwestern Medical Center, Dallas, TX, USA
68. Neuro-X Institute, École Polytechnique Fédérale de Lausanne (EPFL), Lausanne, Switzerland
69. Department of Radiology and Nuclear Medicine, Amsterdam University Medical Center, University of Amsterdam, Amsterdam, The Netherlands
70. Institute of Scientific Instruments of the Czech Academy of Sciences, Brno, Czech Republic
71. Department of Electrical and Software Engineering, Schulich School of Engineering, University of Calgary, Calgary, AB, Canada
72. Center for Advanced MR Development, Department of Radiology, Duke University Medical Center, Durham, NC, USA

73. Department of Diagnostic and Interventional Radiology, University Hospital Ulm, Ulm, Germany
74. Department of Nuclear Medicine, University Hospital Ulm, Ulm, Germany
75. Department of Diagnostic and Interventional Neuroradiology, School of Medicine and Health, TUM Klinikum Rechts der Isar, Technical University of Munich, Munich, Germany
76. TUM-Neuroimaging Center, TUM Klinikum Rechts der Isar, Technical University of Munich, Munich, Germany
77. Psychiatry and Behavioral Neuroscience, Wayne State University School of Medicine, Detroit, MI, USA
78. Athinoula A. Martinos Center for Biomedical Imaging, Institute for Innovation in Imaging, Department of Radiology, Massachusetts General Hospital, Harvard Medical School, Boston, MA, USA
79. Institutionen för experimentell medicinsk vetenskap, Medicinska fakulteten, Lunds universitet, Lund, Sweden
80. Wallenberg Center for Molecular Medicine, Medicinska fakulteten, Lunds universitet, Lund, Sweden
81. Department of Psychiatry, University of California, Los Angeles, Los Angeles, CA, USA
82. Medical Image Analysis Group, Department of Biomedical Engineering, Eindhoven University of Technology, Eindhoven, The Netherlands
83. LIFMET, École Polytechnique Fédérale de Lausanne (EPFL), Lausanne, Switzerland
84. Centre for Human Brain Health and School of Psychology, University of Birmingham, Birmingham, UK
85. Taub Institute for Research on Alzheimer's Disease and the Aging Brain, Columbia University Irving Medical Center, New York, NY, USA
86. CIBM Center for Biomedical Imaging, École Polytechnique Fédérale de Lausanne (EPFL), Lausanne, Switzerland

---

*Corresponding Author
**Name**: Antonia Kaiser
**Email**: antonia.kaiser@epfl.ch/ akaiser.neuro@gmail.com
**Address:**
MR Imaging Technology Section
CIBM EPFL
CH F1 522 Station 6,
1015 Lausanne, Switzerland
+41 21 693 79 69
+41 76 689 90 06

# Graphical Abstract

**Synthetic Data in MR Spectroscopy:**
**Current Practices, Applications, and Considerations**

Synthetic MRS Data Working Group — Led by: J LaMaster, AT Gudmundson, and A Kaiser

- Synthetic data has become essential for MRS, facilitating software validation, deep learning applications, and enhancing reproducibility.
- Synthetic data addresses challenges of data availability, particularly for clinical populations, and offers controlled solutions for investigating uncertainties of in vivo data.
- An overview and considerations for synthetic MRS data is provided.

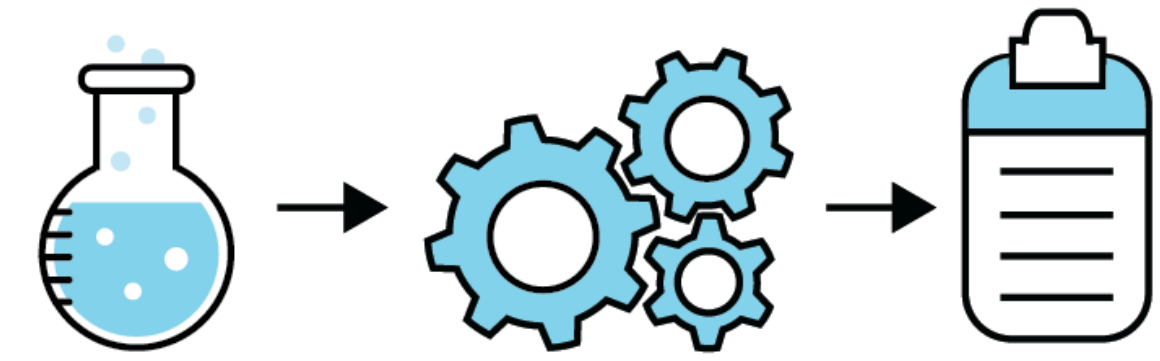

# Abstract/Summary

The use of synthetic data has emerged as an essential tool in Magnetic Resonance Spectroscopy (MRS) research and applications, providing advantages for optimization of acquisition, software validation, deep learning applications, and enhanced reproducibility. Importantly, synthetic data addresses challenges of limited training data availability, particularly for clinical populations, and offers controlled solutions for investigating uncertainties and unexplained variance with *in vivo* data. This work provides a review and evaluation of current practices in the use and generation of synthetic data within the MRS field. Conducted by the MRS Synthetic Data Working Group under the Code & Data Sharing Committee of the MRS Study Group of the International Society for Magnetic Resonance in Medicine (ISMRM), this manuscript encompasses existing literature, supplemented by collective experience and in-house methodologies.

Word Count: 20,418
Figure Count: 7
Table Count: 0 (All 6 tables are in the supplement)

# Abbreviations

2-HG: 2-hydroxyglutarate

AI: Artificial intelligence

Ace: Acetate

Ala: Alanine

Asc: Ascorbate

Asp: Aspartate

bHB: β-Hydroxybutyrate

BOLD: Blood Oxygen Level Dependent

Cho: Choline

Cr: Creatine

CSF: Cerebrospinal Fluid

DL: Deep Learning

dMRS: diffusion-weighted Magnetic Resonance Spectroscopy

EPSI: Echo-Planar Spectroscopic Imaging

fMRS: functional Magnetic Resonance Spectroscopy

GABA: γ-Aminobutyric Acid

Glc: Glucose

Gln: Glutamine

Glu: Glutamate

Gly: Glycine

Glyc: Glycerol

GM: Gray Matter

GPC: Glycerophosphocholine

GSH: Glutathione

IU: Institutional Units

Lac: Lactate

LCM: Linear Combination Model(ing)

ML: Machine Learning

mI: Myo-inositol

MM: Macromolecules

MRS: Magnetic Resonance Spectroscopy

MRSI: Magnetic Resonance Spectroscopic Imaging

NAA: *N*-Acetyl-Aspartate

NAAG: *N*-Acetyl-Aspartyl-Glutamate

NSA: Number of Signal Averages (Transients)

ISMRM: International Society for Magnetic Resonance in Medicine

PCh: Phosphocholine

PCr: Phosphocreatine

PE: Phosphoethanolamine

PRESS: Point Resolved Spectroscopy

RF: Radio frequency

sI: Scyllo-inositol

sLASER: Semi-Localized by Adiabatic Selective Refocusing

SNR: Signal to Noise Ratio

STEAM: Stimulated Echo Acquisition Mode

Tau: Taurine

TE: Echo Time

tCr: Total Creatine (Creatine + Phosphocreatine)

tCho: Total Choline (Glycerophosphocholine + Phosphocholine + Choline)

tNAA: Total *N*-Acetyl-Aspartate (*N*-Acetyl-Aspartate + *N*-Acetyl-Aspartyl-Glutamate)

TR: Repetition Time

WM: White Matter

# 1. Introduction

Magnetic Resonance Spectroscopy (MRS) is a non-invasive technique that enables the measurement of tissue metabolite levels, providing critical insights into metabolic and biochemical changes associated with various diseases (Glunde et al., 2010; Mandal, 2007; Öz et al., 2014). Its applications span both clinical and preclinical research, supporting advancements in neuroscience, oncology, and other medical disciplines.

Despite its utility, MRS faces several challenges that limit its broader adoption and application. Variability in data acquisition methods, limited availability of large datasets, and difficulties in replicating experiments across different sites and scanners hinder the validation and standardization of methods (A. Lin et al., 2021). These challenges slow the progress of the field and delay the development of robust algorithms and reproducible workflows.

One promising approach to address these limitations is the use of **synthetic MRS data**, which involves generating realistic simulated MRS data under controlled conditions. By employing synthetic datasets, researchers can conduct extensive analyses and validations without the ethical or logistical constraints of human or animal experiments. Synthetic data offers a practical and scalable alternative to *in vivo* datasets, overcoming key obstacles related to data acquisition, variability, and accessibility.

Synthetic data provides various critical advantages for MRS research and development. It enables controlled ground truths by allowing precise specification of parameters such as metabolite concentrations, which facilitates rigorous validation of algorithms and models against known values. Additionally, synthetic data is scalable and permits great diversity, as large datasets can be generated economically and computationally efficiently. This scalability supports the optimization of acquisition methods, training of machine learning and deep learning pipelines, and systematic pipeline testing. Synthetic data also supports the study of pathological cases by using ground truth metabolic profiles derived from rare or extreme *in vivo* conditions to generate data across a wide range of acquisition settings. This approach enables systematic evaluation and optimization of measurement protocols and analysis methods under conditions that are difficult to sample empirically. Furthermore, synthetic datasets are valuable for teaching MRS techniques and validating methods under well-defined conditions, providing clear benchmarks for reproducibility. Finally, in clinical settings where data availability is limited and inter-subject variability is high, synthetic datasets allow controlled testing of algorithm performance under clinically relevant variability, supporting assessment of stability, bias, and transferability across heterogeneous populations. By addressing these needs, synthetic data play an indispensable role in overcoming the challenges associated with MRS research and applications.

While synthetic data is capable of closing many gaps in MRS research, the field is still in the early days of development of generation methodologies. For many researchers, simulating a realistic MRS dataset can be an overwhelming task with no clear starting point. For the field at large, the definition of what is “realistic” and appropriate, is ambiguous and unregulated. Such definitions change depending on the targeted application and evolve with the development of new methodologies and empirical findings.

To provide clarity and guidance for the field, this manuscript outlines existing works on generating, using, and leveraging synthetic MRS data. We begin by detailing the **core components** necessary for constructing the essentials and foundation of synthetic spectra. Next, the **advanced components** are presented in an in-depth overview of the next level of features that allow researchers to simulate specialized or non-standard scenarios. We then expand to **modality-specific considerations**, highlighting how synthetic data can be adapted for different MRS techniques and nuclei. Subsequently, we describe **application contexts**, illustrating how synthetic datasets can support a wide range of research questions, tool development, and methodological benchmarking. Finally, we present current **validation approaches** for assessing synthetic data quality and realism and propose **standardized file formats and general reporting guidelines** to promote reproducibility, interoperability, and transparent dissemination within the MRS community.

This review, conducted by the MRS Synthetic Data Working Group from the Code & Data Sharing Committee of the MRS Study Group of the International Society for Magnetic Resonance in Medicine (ISMRM), includes insights from the literature and contributions from active researchers (please find a list of publications in Supplementary Table S1 Master Table). By identifying gaps and opportunities, this work aims to guide researchers in using synthetic data effectively to advance MRS research and applications. The preparation of this manuscript has also provided an important opportunity to develop considerations and highlight where differences lie.

In each section of this manuscript, we begin by outlining the 'Considerations' necessary for robust synthetic MRS data generation and applications (please find a summary of these in the Supplements S2), followed by a detailed discussion of current practices. A comprehensive master table of publications (Table S1 Master Table) employing synthetic data in these contexts accompanies this manuscript and will remain a living document that can be updated by the community as new contributions emerge.

The ‘Considerations’ in this work are **not prescriptive requirements**, but rather useful suggestions. There are many choices to make when simulating synthetic MRS data, and these choices merit careful consideration and discussion in future publications. Our working group acknowledges that deviation from these suggestions *can be appropriate* in various scenarios, but we believe that justification should always be provided.

# 2. Core Components for Synthetic MRS Data

## 2.1. Basis Sets

***Considerations:***
***General:***

- *Synthetic data should be derived from a defined basis set (broadly including simulated or experimentally acquired basis functions). The basis set should be provided alongside the synthetic dataset.*
- *Essential acquisition parameters (field strength, sequence type, echo time [TE], spectral width [SW], and number of data points) should be documented in full.*
- *J-coupling evolution and accurate metabolite basis functions should be incorporated into the signal model.*

***$^{1}$H MRS:***

- *Signal models should reflect common in vivo scenarios relevant to the tissue under investigation (e.g., human brain).*
- *For echo times below approximately 80 ms, inclusion of macromolecular (MM) signals is strongly encouraged.*

Basis sets are a foundational component in MRS data simulation and analysis, enabling the deconvolution of complex spectral data into meaningful biochemical information. These sets mathematically represent individual metabolite signals and are essential for linear combination modeling (LC modeling) of MR spectra. While the development and validation of basis sets have been extensively reviewed in prior works (Near et al., 2020; Provencher, 1993), some essential points must be reiterated due to their critical importance in synthetic data simulations.

The quality and compatibility of a basis set with the experimental setup directly impact the accuracy of MRS quantification. Basis sets are typically generated under specific acquisition parameters, such as for a given field strength, specific pulse sequence, echo time (TE), spectral width (SW), and the number of data points, using simulations or experimental data thereof. These parameters capture the nuances of metabolite behavior, including chemical shifts and coupling patterns, which are crucial for realistic synthetic MRS data. Basis set composition is a critical determinant of fitting performance when analyzing experimentally acquired spectra that closely resemble the intended acquisition protocol. In a synthetic data context, careful consideration is required, as the selected metabolite and macromolecular components define the model space and may bias validation or benchmarking outcomes. Several studies have systematically investigated how basis set composition affects quantification accuracy and robustness (Hofmann et al., 2002, Lazariev et al., 2011, Davies-Jenkins et al., 2026, Emeliyanova et al., 2026). Although a comprehensive evaluation of basis set optimization is beyond the scope of this manuscript, readers interested in detailed metabolite-specific physical properties can consult the living MRS-Sim database, which compiles moiety-specific spin definitions and relevant simulation parameters; MRS-Sim GitHub repository (LaMaster et al., 2025).

Metabolite basis sets are often derived from:

- **Mathematical models** (elaborated on below);
- **Phantom experiments;**
- **Density-matrix simulations,** which account for chemical shifts and couplings under specific pulse sequence conditions (Farrar, 1990).

Various dedicated software packages can generate these basis sets, including (see a detailed table in Supplements; Supplementary Table S2):

- **GAMMA/PyGAMMA** (C++; Python): Offers high flexibility for quantum mechanical simulations (Smith et al., 1994);
- **MARSS** (MATLAB): User-friendly, fast, and optimized for typical MRS studies (Landheer et al., 2021);
- **FID-A** (MATLAB): Comprehensive and open-source (Simpson et al., 2017);
- **Vespa** (Python): Flexible and well-documented for advanced users (Soher et al., 2023);
- **MRSCloud** (Web-based): Accessible to users with limited local resources (Hui et al., 2022);
- **NMRScopeB** (Java; Python): seamlessly integrates advanced NMR analysis (Starčuk & Starčuková, 2017).

These packages provide predefined metabolite spin definitions, common pulse sequences, and adjustable parameters, ensuring compatibility with various experimental setups. Additionally, presimulated basis sets can be found on several publicly accessible databases (https://mrshub.org/datasets_basissets/).

### 2.1.2. Macromolecular Basis Sets

Macromolecular (MM) signals, representing broad resonances in MR spectra, are typically modeled using (Supplementary Table S3 contains a more comprehensive list of literature):

- Parameterized mathematical functions (e.g., Gaussian or Voigtian resonances).
- Basis functions derived from *in vivo* experiments, such as metabolite-nulled spectra matched for field strength and TE (Cudalbu et al., 2021).

Some synthetic datasets (Oeltzschner et al., 2020; Xiao et al., 2024; Zöllner et al., 2024) include sequence-specific MM basis functions, either as raw experimental data or as cleaned, noise-free models (Marjańska et al., 2022). Resources like MRSHub (https://mrshub.org/datasets_mm/) provide growing libraries of MM experimental data for broader use.

### 2.1.3. X-nuclei Basis Sets

For X-nuclei MRS (e.g., $^{31}$P, $^{13}$C, $^{23}$Na), basis sets often rely on individual lineshape functions constrained by prior knowledge of parameters like chemical shifts and coupling constants. Tools such as AMARES (Vanhamme et al., 1997; Xu et al., 2024), OXSA (Purvis et al., 2017), and jMRUI (Stefan et al., 2009) facilitate the creation and simulation of these datasets. Additionally, packages like LCModel (Provencher, 2001), FSL-MRS (Clarke et al., 2021) and MARSS (Landheer et al., 2025) support simulation of X-nuclei basis sets for both research and clinical applications.

While some $^{31}$P datasets include field-specific parameters for metabolites like NAD$^+$/NADH (Deelchand, Nguyen, et al., 2015a; Lu et al., 2014), other nuclei such as $^{23}$Na and $^{13}$C remain underexplored in synthetic simulations, presenting opportunities for future development. For $^{2}$H MRS, spin system parameters are largely unchanged when J-couplings are scaled by the respective gyromagnetic ratios and the simulation program can properly accommodate spin 1 nuclear evolution (Landheer et al., 2025).

## 2.2. Signal Models

***Considerations:***

- ***Noise:*** *Simulated datasets should span coil-combined signal-to-noise ratios (SNRs) representative of typical in vivo acquisitions (approximately 5–75), while allowing extension beyond this range when required for specific benchmarking applications.*
- ***Linebroadening:*** *Both Lorentzian and non-Lorentzian broadening mechanisms should be implemented to approximate realistic in vivo spectral lineshapes.*
- ***Phase and Frequency Shifts:*** *Global phase and frequency shifts should be incorporated into the final synthetic spectrum to enhance realism. Small metabolite-specific frequency shifts may be included where appropriate.*
- ***Macromolecular**, **lipid**, and **baseline** components should be modeled where relevant to the intended in vivo scenario.*

Similar to linear combination-based modeling approaches, synthetic data applies a signal model to a set of metabolite basis functions. This involves modifying several free parameters to adjust the basis functions, which are then combined into a plausible approximation of an acquired spectrum. At minimum, a signal model should include amplitude and lineshape adjustments to integrate the basis set into a single spectrum. However, additional signal components would need to be implemented to better mimic *in vivo* data and its artifacts (Bugler et al., 2025). A general overview of a simulation workflow is shown in Figure 1.

### 2.2.1. Amplitude

Metabolite-specific amplitude parameters are used to linearly scale basis functions. These values are often derived from distributions informed by large datasets or population-specific priors (Gudmundson et al., 2023; Gudmundson et al., 2023a; Oeltzschner et al., 2020; Xiao et al., 2024; Zöllner et al., 2024). Simulators can extend this by incorporating tissue-specific distributions (Gudmundson et al., 2023; Gudmundson et al., 2023a; Oeltzschner et al., 2020; Zöllner et al., 2024), or dynamic changes, such as those in $T_2$ series (Zöllner et al., 2024) or task-related fMRS experiments (Craven et al., 2024). Recent simulation frameworks, such as MRS-Sim, have introduced modular approaches that allow for the flexible integration of these amplitude parameters, enhancing the realism and adaptability of synthetic MRS data generation (LaMaster et al., 2025).

### 2.2.2 Phase

Zero- and first-order phase shifts applied to the final synthetic spectrum are typically drawn from zero-centered distributions to reflect *in vivo* variability (Borbath et al., 2021; Gudmundson et al., 2023; LaMaster et al., 2025; Oeltzschner et al., 2020; Rizzo et al., 2023; Zöllner et al., 2024,

Bugler et al., 2025). While common, their inclusion is not universal (Craven et al., 2024; Simicic et al., 2024).

### 2.2.3 Frequency Shift

Global frequency shifts can be applied to the final synthetic spectrum to simulate experimental $B_0$ changes caused by thermal drift, subject motion, etc. Small metabolite-specific frequency shifts are applied to account for factors like temperature and pH variations and spin-system inaccuracies. These shifts are often drawn from random distributions (Borbath et al., 2021; Gudmundson, Davies-Jenkins, et al., 2023; LaMaster et al., 2025; Oeltzschner et al., 2020; Rizzo et al., 2023; Zöllner et al., 2024, Bugler et al., 2025) or obtained from measured *in vivo* $B_0$ maps (Motyka et al., 2019).

### 2.2.4 Lineshape

**Non-Lorentzian:** Non-Lorentzian line broadening arises from local $B_0$ inhomogeneities, e.g., local field susceptibility and imperfect shimming. Gaussian lineshapes can be used to describe the effect of normally distributed intra-voxel field inhomogeneities (Borbath et al., 2021; Craven et al., 2024; Gudmundson et al., 2023; LaMaster et al., 2025; Marjańska et al., 2022; Rizzo et al., 2023; Zöllner et al., 2024). More complex definable lineshape convolution kernels or simulations including measured $B_0$ maps (Motyka et al., 2019) can be used to describe non-normality of $B_0$ within the voxel, and while this can be used in simulations, to date, it has only been used in spectral fitting routines (Oeltzschner et al., 2020; Provencher, 2001; Slotboom et al., 1998). Both are usually applied equally to all basis functions, although some simulators differentiate between MMs and metabolites (Borbath et al., n.d.; LaMaster et al., 2025), and one added a per-metabolite jitter (Gudmundson et al., 2023).

**Lorentzian:** Most, but not all (Simicic et al., 2024), simulators apply global (Marjańska et al., 2022), metabolite-specific (Borbath et al., 2021; Craven et al., 2024; Gudmundson et al., 2023; Oeltzschner et al., 2020; Xiao et al., 2024; Zöllner et al., 2024), or even moiety-specific (Gudmundson et al., 2023; Rizzo et al., 2023) Lorentzian line broadening to mimic the effects of $T_2$ relaxation times on the lineshape. When available and applicable, tissue-specific $T_2$ literature or measured values are used (Oeltzschner et al., 2020; Rizzo et al., 2023; Zöllner et al., 2024) to account for tissue-specific relaxation. They may also be modulated across a time series of spectra to simulate BOLD effects in fMRS experiments (Craven et al., 2024), as elaborated on in the fMRS section.

**Voigtian:** The Voigt function describes a signal decay resulting from the multiplication of exponential and Gaussian decays in the time domain; in the frequency domain, this corresponds to a convolution of Lorentzian and Gaussian lineshapes. It represents the combined effects of homogeneous broadening (e.g., relaxation-related exponential decay processes contributing to $T_2$*) and inhomogeneous broadening arising from normally distributed intra-voxel magnetic field variations. The Voigt profile is the common lineshape encountered in *in vivo* data (Marshall et al., 1997).

## 2.2.5 Noise

Most simulators and synthetic datasets include uncorrelated Gaussian noise along with the final metabolite spectrum (Borbath et al., 2021; Gudmundson et al., 2023; Marjańska et al., 2022; Mosso et al., 2022; Oeltzschner et al., 2020; Rizzo et al., 2023; Simicic et al., 2021; Xiao et al., 2024; Zöllner et al., 2024, Bugler et al., 2025); some can add noise correlations between transients or coils (LaMaster et al., 2025; Oeltzschner et al., 2020; Zöllner et al., 2024). To simulate realistic acquisition conditions, the noise is often sampled independently in the real and imaginary parts and scaled to achieve a desired signal-to-noise ratio (SNR). Definitions of target SNR are generally well-defined within a given work, but the definitions can vary considerably and therefore are difficult to compare across studies.

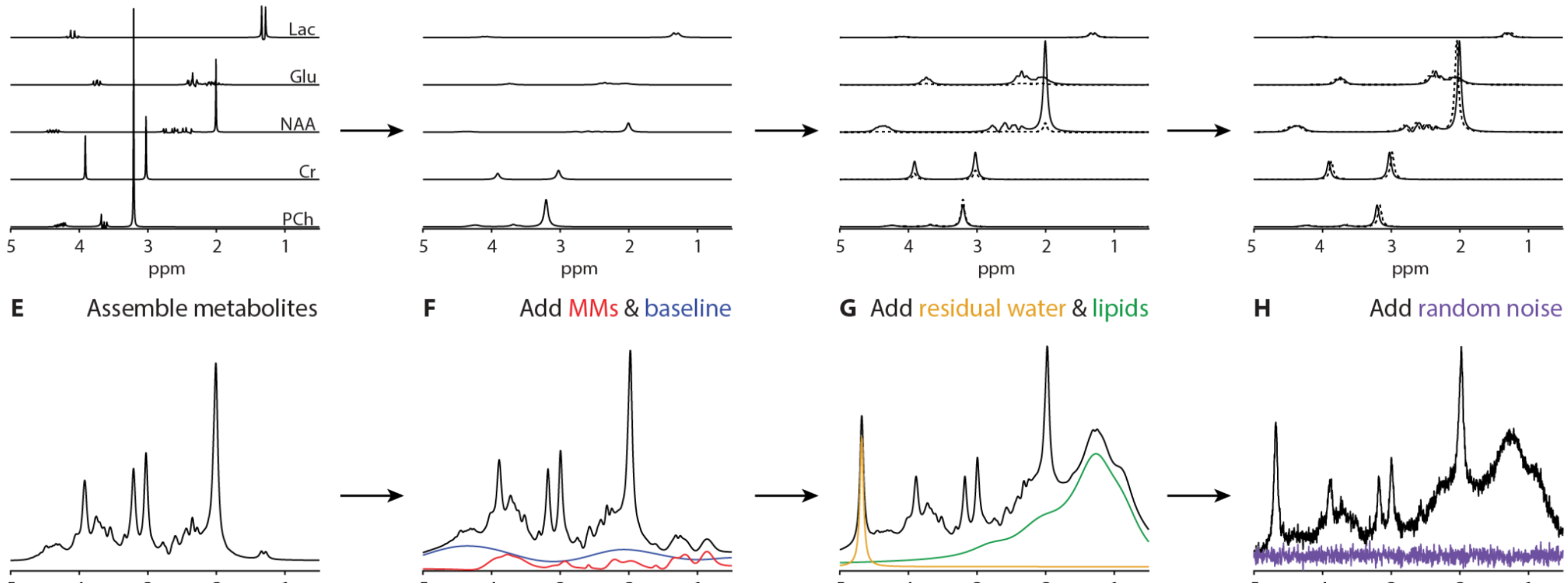

**Figure 1. Workflow for generating an in vivo–like synthetic MRS spectrum.** Schematic overview of the sequential steps involved in constructing a realistic synthetic spectrum: **A)** simulation of a metabolite basis set with the desired pulse sequence; **B–D)** application of line broadening, amplitude scaling, and frequency shifts; **E)** summation of metabolite components; **F–H)** addition of macromolecular signal, baseline distortions, residual water/lipid contributions, and random noise to approximate in vivo spectral characteristics.

## 2.2.6 X-Nuclei

Synthetic data generation for X-nuclei MRS differs from $^{1}$H MRS. Notably, water suppression, lipid removal, and MM contributions are mostly unnecessary to model, as these components do not dominate the spectra of most X-nuclei. Because the gyromagnetic ratio of X-nuclei is lower than that of $^{1}$H, both their linewidths (from $B_0$ inhomogeneity) and their signal amplitudes are reduced. Simulations should scale these parameters accordingly, even for voxel sizes bigger than those of $^{1}$H acquisitions. Furthermore, the look-and-feel of a typical *in vivo* X-nuclei MRS spectrum can vary dramatically across field strength due to the dipolar broadening effect being more prominent at lower field strengths (Jensen et al., 2002), acquiring data with or without $^{1}$H-decoupling at low fields (Potwarka et al., 1999), and the pre-acquisition delay time (Stanley, 2000). For $^{31}$P-MRS specifically, synthetic data should account for: frequency shifts arising from variations in intracellular pH and magnesium ([$Mg^{2+}$]) concentrations (Franke et al., 2024; Korzowski et al., 2020); homonuclear J-coupling, particularly in ATP, to reproduce the characteristic multiplet structures (Deelchand, Nguyen, et al., 2015b; Lu et al., 2016); and effects of magnetization transfer in dynamic studies that result from enzymatic activity (Du et al., 2006; Ren et al,. 2017; Widmaier et al., 2025). These factors play a critical role in the estimation of the membrane phospholipid and high-energy phosphate metabolite levels. It is also essential to understand that most X-nuclear MRS is dynamic in nature (e.g., $^{2}H$ (Osburg et al., 2025, Duguid et al., 2025), $^{31}P$ (García et al., 2025, Songeon et al., 2025), HP-$^{13}C$ (Brindle et al., 2011, Li et al., 2021)), which plays a critical role for realistic synthesis (see Section 3.4).

## 2.3 Metabolites: Concentration and Relaxation Ranges

***Considerations:***

- ***Type:*** *The simulation framework should explicitly state whether parameters are defined as fixed single values or sampled from a distribution.*
- ***References:*** *All parameter ranges should be justified and supported by appropriate literature references or empirical data.*
- ***Choice*****:** *Parameter ranges should be selected according to the intended purpose of the simulation. For software validation and benchmarking, ranges should be sufficiently broad to challenge fitting stability, robustness, and sensitivity. For simulations designed to reflect clinical scenarios, parameter limits should prioritize physiological plausibility and expected patient variability.*
- ***Implementation:*** *It should be explicitly acknowledged that metabolite concentrations and spectral amplitudes are not equivalent quantities. The relationship between concentration and signal amplitude (including sequence timing, relaxation effects, and scaling factors) should be clearly defined within the simulation model.*

Metabolite ranges in synthetic MRS data refer to the minimum and maximum values used to define a distribution of metabolite concentration levels and relaxation rates, corresponding to signal amplitudes and lineshapes, respectively. Accurate metabolite ranges are essential when generating synthetic MRS data, as the development of software packages for processing and quantification relies upon synthetic data serving as a ground truth reference and being representative of the *in vivo* target data. Therefore, it is desirable to use ranges that are consistent with those present in biological tissue and also consider that these values are greatly influenced by organ, anatomical brain regions (Pouwels & Frahm, 1998), tissue type (Harris et al., 2015), age (Angelie et al., 2001; Gong et al., 2022), sex/gender, metabolic state (e.g., fasting vs. post-prandial), hormonal status (e.g., menstrual cycle, menopause) (Chang et al., 2009; Cichocka et al., 2018; Maddock et al., 2006, Song et al, 2025), and of course pathology. Additionally, non-biological factors such as scanner vendor (Bell et al., 2022; Považan et al., 2020) and acquisition sequence (Bell et al., 2025a; Bell et al., 2025b; Hancu, 2009) have been shown to influence quantified metabolite values and may need to be reflected when defining synthetic ranges for specific use cases.

Metabolite concentrations in the literature are reported in various units (Gudmundson et al., 2023a), including ratios relative to water or reference metabolites (commonly tCr), 'absolute' units such as mol/kg or mol/L, and institutional units (IU). ‘Absolute’ quantification requires correction for relaxation effects and appropriate signal scaling relative to a known reference (e.g., internal water, external phantom, or electronic reference methods) (Near et al., 2020), synthetic MRS studies often adopt whichever reporting convention aligns with their simulation strategy. Transparency about the chosen units and how they map onto simulated amplitudes is essential, as this determines how concentration ranges are interpreted during synthetic data construction.

### 2.3.1. Concentrations

Metabolite concentration ranges for generating synthetic MRS data can be determined through employing meta-analytic methods across a collection of *in vivo* research literature, performing single study-specific or hypothesis-driven experiments, or a combination of both. Literature-derived values provide broad coverage of typical metabolite levels across tissues, conditions, and populations, enabling researchers to define general realistic ranges (Gudmundson et al., 2023a). Experimental measurements, on the other hand, offer flexibility for targeting specific research questions or simulating conditions that are underrepresented or insufficiently characterized in the literature. Together, these approaches allow concentration ranges to be tailored to the intended biological context of the synthetic dataset.

To examine how synthetic MRS data is typically constructed, we conducted a focused literature review of metabolite concentration values used in simulations. A total of 24 studies employing synthetic MRS data were identified (see Supplementary Information section 3). Comparability across studies was limited by a mixture of reporting units and heterogeneous levels of methodological detail. Eight studies reported concentrations in millimolar (mM), the most common unit among the surveyed literature, and these were therefore used for aggregated visualization. Figure 2 shows the mean and range of each metabolite based on the mM values reported.

Across these studies, the concentration ranges expressed in mM were relatively consistent and generally reflected values typical of healthy brain tissue, unsurprising given that synthetic datasets are most frequently used for algorithm or pipeline validation (de Graaf, 2019; Govindaraju et al., 2000)(see also Section 5.3). When comparing these simulation-derived ranges to values reported in recent meta-analyses (Gudmundson et al., 2023a), the central tendency aligns reasonably well (e.g., Glx ≈ 11.5 ± 4.2 mM in the meta-analysis). However, the reported ranges used in the simulations tended to be narrower. This likely reflects both the limited number of simulation studies (N = 8 reporting in mM) and the fact that most simulations target “typical healthy” concentrations rather than attempting to represent the full variability reported in the literature, including values from larger samples or heterogeneous clinical populations.

These observations highlight the importance of clearly defining and justifying the intended physiological scope of synthetic metabolite ranges, whether they are meant to represent typical healthy values, narrower application-specific ranges, or broader variability for algorithm stress-testing.

2.3.1.1. Amplitudes from Concentrations

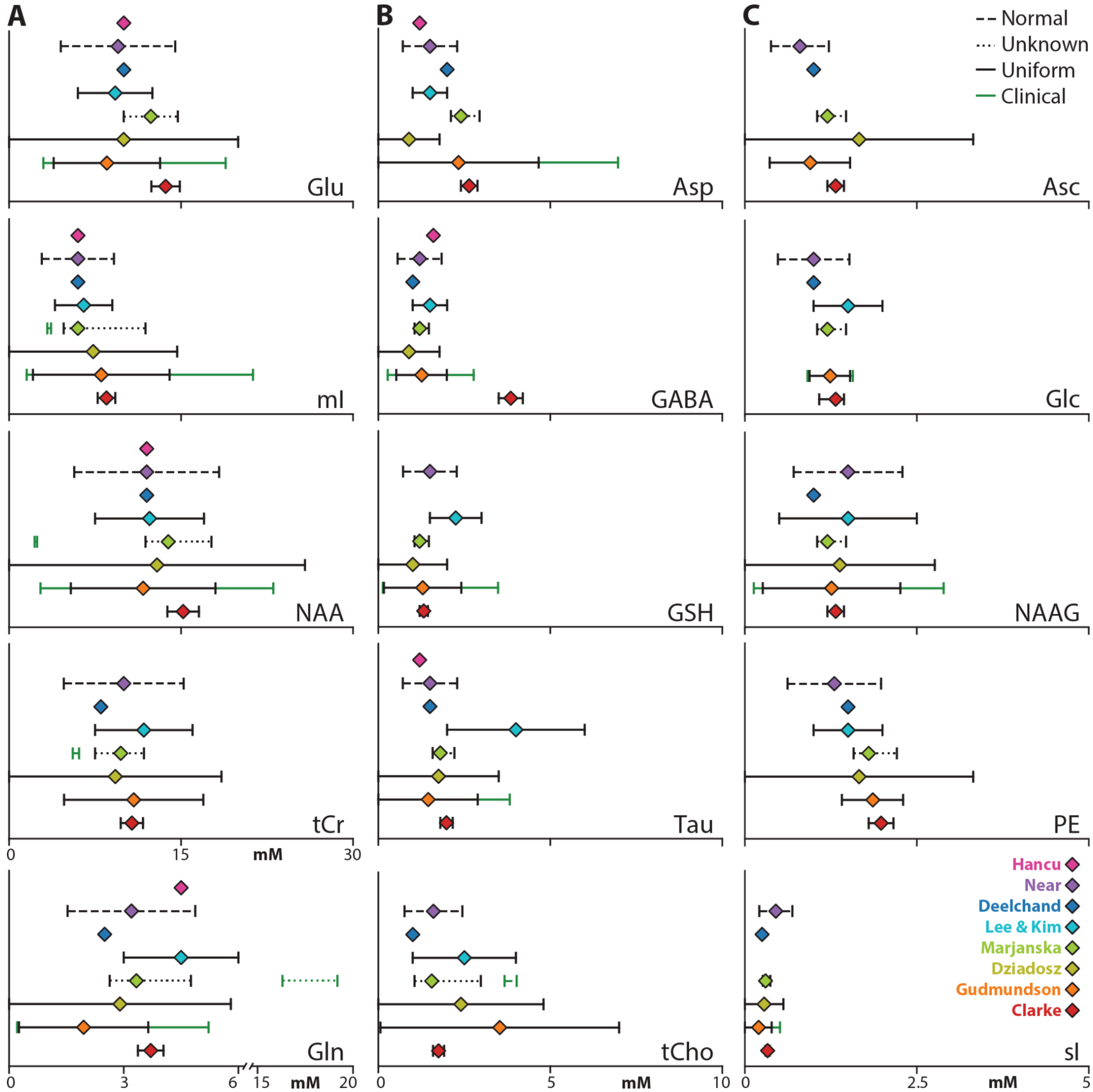

**Figure 2**. Concentrations for **A)** high (0-30 mM), **B)** medium (0-10 mM), and **C)** low (0-5 mM) concentrated metabolites used in synthetic MRS data. Diamond markers indicate the mean concentration value. The statistical distribution can be seen as a dashed line for normal, solid line for uniform, dotted line for unknown distributions, when a range was provided. Green lines were given to datasets that provided an additional clinical range with the exception of Dziadosz et al., 2023 whose values for healthy and clinical populations could not be separated.

Metabolite concentrations are used to determine the amplitude coefficients that scale each basis function during synthetic spectrum generation. In practice, most simulation frameworks do not attempt to model the full physical signal scaling that would link in vivo concentrations to measured signal amplitudes. Such a mapping would require explicit simulation of the unsuppressed water reference signal, together with assumptions about tissue composition, relaxation properties, and water visibility (Near et al., 2021). Instead, synthetic MRS studies typically use concentration ratios to specify relative amplitudes. This is commonly implemented in one of two ways.

The first approach selects a reference metabolite, historically tCr, normalizes all basis functions to this reference, and then applies literature-derived concentration ratios to determine their relative amplitudes (Govindaraju et al., 2000; Graaf, 2019). This method preserves realistic metabolite relationships without requiring 'absolute' quantification. The second approach assumes that each basis function corresponds to the signal produced by a 1.0 mM solution of that metabolite, which allows millimolar literature concentrations to be used directly as amplitude scaling factors (Clarke et al., 2021; Soher et al., 2023). In both cases, the resulting amplitude coefficients represent relative signal levels rather than 'absolute' concentrations, a behavior reflected in several modern simulation and analysis toolboxes (*LCModel User's Manual*, n.d.; Oeltzschner et al., 2020).

Transparent reporting of how amplitudes were derived, including the assumed relationship between concentration values and basis function scaling, is essential. Clear documentation ensures that synthetic data can be reproduced, interpreted correctly, and compared across studies, and it provides necessary context for downstream quantification or algorithm validation.

### 2.3.2. Relaxation

Relaxation properties contribute critically to the appearance of synthetic MRS data because they influence both signal amplitude and lineshape. Although the studies reviewed were relatively consistent in their stated choices of metabolite concentrations, there were notable differences in how relaxation was implemented. Some studies applied metabolite-specific relaxation constants, using measured or literature-derived $T_1$ and $T_2$ values for individual metabolites or metabolite moieties (e.g., Ernst et al., 1993; Mlynárik et al., 2001; Deelchand et al., 2010). Others scaled all metabolite amplitudes by a fixed relaxation factor relative to water, which simplifies implementation but does not capture differences in relaxation across metabolites (e.g., Hancu, 2009; Clarke et al., 2021). A smaller number of studies applied no relaxation-based modulation of amplitude or linewidth, producing spectra that may not reflect realistic *in vivo* characteristics.

These different implementations can result in synthetic spectra with markedly different appearances. $T_1$ relaxation determines longitudinal recovery during the acquisition sequence and has a substantial effect on signal amplitude, particularly in short TR experiments. $T_2^*$ relaxation governs transverse decay and contributes to Lorentzian broadening, thereby influencing spectral resolution and the practical visibility of J-resolved multiplet structure. Because metabolite $T_1$ and $T_2$ values vary across field strength, tissue type, and acquisition parameters  (Stanley et al., 1995; Marjańska et al., 2012; Wansapura et al., 1999), the choice of relaxation model should reflect the intended simulation context.

## 2.4. Challenges and next steps

### 2.4.1. Basis Sets

The development and application of basis sets are central to generating realistic synthetic MRS data. Advances in tools, improvements to spin system definitions, and the incorporation of real pulse and gradient shapes that reflect the sequence used on the scanner have substantially increased the fidelity of simulated basis sets. Despite these advances, several challenges remain.

Variability in sequence parameters, particularly pulse timings, and the widespread use of ideal radiofrequency pulses during basis set simulation remain persistent sources of discrepancy. Many real-world sequences rely on proprietary or undocumented pulse shapes, making it difficult to simulate them with perfect fidelity. Sequence imperfections, including those arising from hardware limitations such as gradient nonlinearity or $B_1$ inhomogeneity, are also frequently overlooked.

Macromolecule, lipid, and baseline signal simulation vary considerably across synthetic data generators, and these differences can meaningfully affect linear-combination modeling results. Since the impact of broad signal modeling on quantitative fitting is well documented (Cudalbu et al., 2021; Kreis, 2016, p. 201), careful evaluation is needed to determine whether particular simulation frameworks unintentionally favor certain modeling algorithms.

A promising idea to increase flexibility is to oversample the temporal sampling grid during basis set simulation, which allows a single high-resolution simulation to be downsampled to multiple commonly used spectral widths by subsampling the time points. Further exploration of this approach may reduce the need to regenerate basis sets when acquisition parameters vary within typical ranges (Gudmundson et al., 2023).

### 2.4.2. Signal Models

Synthetic data generators vary widely in how they implement key components of the signal model, including eddy currents, spurious echoes, out-of-voxel contamination, residual water, motion artifacts, and other nuisance signals. Without sufficient methodological detail, this variability limits the ability to compare synthetic datasets and complicates reproducibility across platforms (see Section 3).

Even basic definitions can differ. For example, the consensus definition of SNR from the Terminology Consensus (Kreis et al., 2021) is not consistently adopted across existing tools. This variation reflects differences in the intended complexity of simulation frameworks, which range from simple weighted combinations of ideal basis functions to detailed reproductions of pulse sequence physics in spatially varying geometries with accurate timing information. Adoption of the consensus SNR definition is strongly encouraged to facilitate comparison, reproducibility, and transparency.

A significant next step is the ability to synthesize a corresponding unsuppressed water reference signal (Kyathanahally & Kreis, 2017). Generating an unsuppressed water signal would support several goals. First, it would allow metabolite- and water-specific $T_1$ and $T_2$ relaxation times and diffusion properties to be incorporated into amplitude and linewidth modeling consistently. Second, it would align synthetic datasets more closely with *in vivo* quantification workflows, which typically rely on water scaling for amplitude normalization. Third, it would enable systematic assessment of fitting accuracy in 'absolute' quantification pipelines, thereby bridging a long-standing gap in validation efforts.

In parallel, incorporating realistic tissue fraction data is essential because metabolite concentrations differ across gray matter, white matter, and cerebrospinal fluid. Accounting for tissue composition and effective water visibility (i.e., compartment- and relaxation-dependent differences in observable water signal across GM, WM, CSF, and vascular contributions) improves the physiological realism of synthetic spectra, particularly when generating synthetic unsuppressed water references.

### 2.4.3. Metabolite Ranges

Defining metabolite concentration ranges that are suitable for all use cases remains challenging. Users must remain aware that narrowly chosen ranges may not capture the variability observed across different acquisition methods, anatomical regions, age groups, field strengths, and scanner vendors. Broader literature sources or meta-analyses, such as Gudmundson et al., 2023a, may provide appropriate guidance when natural variability is substantial. Conversely, when simulating specific scenarios, for example, particular brain regions or clinical cohorts, ranges should be tailored to reflect the relevant physiological context.

A major next step involves clarifying not only the choice of ranges but also the distributions from which synthetic concentrations are drawn. Specifying minimum and maximum values often leads users to apply a uniform distribution by default, which may not reflect the true variability of metabolite concentrations *in vivo*. More appropriate alternatives may include normal distributions, skewed Gaussian distributions, or Poisson-like distributions, depending on the metabolite and target population. Clear recommendations regarding distribution shape would help users generate synthetic datasets that more accurately represent underlying biological variability.

Our review did not evaluate macromolecule and lipid concentrations because their amplitudes are sparsely reported and span broad ranges. These resonances are frequently modeled separately from metabolites, creating uncertainty in their relative amplitude scaling. As macromolecular signals significantly influence quantification, improved guidance for MM and lipid amplitude modeling is warranted.

Finally, until simulations routinely incorporate synthetic unsuppressed water references, bridging the gap between metabolite concentrations and the amplitude values used in simulations remains an open challenge. Detailed reporting of the choices made, including the selection of ranges, the type of distribution applied, and the strategy for mapping concentration to amplitude are essential for reproducibility and accurate interpretation.

# 3. Advanced Components for Synthetic MRS Data

While core components establish the bare minimum for synthetic MRS data (see *previous section*), advanced components are essential for maximal realism. This section reviews features included in the simulation of synthetic datasets that go beyond minimum considerations, incorporating additional signal components to better reflect real-world MRS data.

Advanced components address the intricate interplay between spectral features and experimental conditions. For instance, the spectral dimension for simulated metabolites and macromolecules must account for chemical shifts, J-coupling, and their interactions with water and lipid-suppression modules, RF pulses, $B_1^+$ field inhomogeneities, $B_0$-field variations, and $T_1$ and $T_2$ relaxation times. But simulating these complex interactions is computationally intensive, which has historically limited their incorporation in synthetic data models. Most existing simulations simplify these factors, focusing only on chemical shifts, J-coupling, or singlet resonances. However, as computational resources and modeling techniques advance, integrating these sophisticated interactions becomes increasingly feasible, offering more accurate and application-relevant synthetic datasets.

## 3.1. Signal Components

***Considerations:***

- ***1-H:***
  - ***Macromolecular** signals should be included with tissue-specific scaling appropriate to the simulated anatomical or physiological context.*
  - ***Baseline** contributions should be modeled using flexible approaches such as cubic splines, Voigt lines, or polynomial functions, depending on the intended in vivo scenario and fitting strategy being evaluated.*
  - *Common **nuisance** signals, including residual water, lipids, and echo artifacts, should be incorporated where relevant to the acquisition type and study objective.*

### 3.1.1 Baseline

According to recent terminology consensus (Kreis et al., 2021), ‘baseline’ describes broad nuisance signals underlying the metabolite spectra of interest (e.g., from poor water suppression), but not MM. Generally, some (but not all (Marjańska et al., 2022; Rizzo et al., 2023)) synthetic data generators add broad signals to achieve this, but while their parameterization varies, many choose to generate baselines using the same methods they use for modeling them. Cubic splines are a common choice (Borbath et al., 2021; Oeltzschner et al., 2020; Xiao et al., 2024; Zöllner et al., 2024, Bugler et al., 2025), as are very broad Voigtian signals (Simicic et al., 2021) and Gaussian peaks (Wilson, 2021). However, some work has explored unconventional approaches, such as constructing the baseline from *in vivo* fit residuals (Craven et al., 2024) or using a model-agnostic smoothed, pseudo-random walk algorithm (LaMaster et al., 2025).

### 3.1.2 Nuisance Signals

*In vivo* spectra further contain partially suppressed water, lipids, and out-of-voxel or “ghost” signals. These nuisance signals are challenging to parametrize and therefore less commonly included in synthetic data and usually included when they are the specific focus.

**Residual Water:** Water artifacts from partially suppressed water are simulated either by considering sequence parameters, relaxation weighting and variable water suppression due to off-resonance and flip angle errors or simply by adding Lorentzian lines of variable amplitude but fixed chemical shift. Some simulators describe residual water with a single simulated peak (Marjańska et al., 2022)(with varying lineshape, frequency, width, phase, and amplitude) or multiple peaks (Gudmundson et al., 2023) to mimic the resemblance of imperfect *in vivo* water suppression, while another used a smoothed pseudo-random walk algorithm to generate random signal fluctuations for the residual water region (LaMaster et al., 2025).

**Lipids:** Artifacts from extracranial lipids may be modeled randomly or based on *in vivo* MRSI data to consider natural variability of lipid amplitudes and patterns or by simulating singlets at known chemical shifts. Lipid contamination is incorporated in some simulators by optionally adding one or more Voigt lines with defined center frequencies, widths, and amplitude distributions (Oeltzschner et al., 2020; Rizzo et al., 2023; Zöllner et al., 2024, Bugler et al., 2025) or extracted from high-lipid *in vivo* data (Marjańska et al., 2022; Weiser et al., 2025). An alternative modeling approach is the use of, e.g., cubic basis splines (Wilson, 2021).

**Spurious Echoes:** Some simulators have explicit models of spurious echoes (Gudmundson et al., 2023). One simulator parametrizes the spurious echo as a complex time-domain signal by a center time point, width, frequency, zero-order phase, and amplitude (Gudmundson et al., 2023, Bugler et al., 2025). Since spurious echoes arise from unwanted coherence pathways, a more physically grounded approach is to model these pathways directly. The coherence-pathway origin of spurious signals and their sequence-dependent behavior is discussed in Soher et al., 2023.

## 3.2. Spatial Components

**_Considerations:_**

- ***Coil-Specific Factors***: *Array coil data should be simulated with realistic noise correlations, receive sensitivity profiles, and head-position variability to approximate in vivo acquisition conditions.*
- ***Spatial Field Variations***: *Spatial inhomogeneities in $B_0$ and $B_1$ fields, as well as spatially varying metabolite distributions, should be incorporated into the simulation framework. This is particularly important for MRSI datasets, where voxel-specific variations should reflect realistic in vivo patterns.*
- ***k-Space Sampling***: *Realistic k-space encoding schemes should be implemented, including conventional phase encoding and alternative trajectory-based spatial–spectral encoding techniques. Gradient imperfections and system-related distortions should be modeled where feasible to improve the accuracy of the simulated spatial response function.*

Synthetic MRS data can be made substantially more realistic by incorporating spatially dependent factors that influence the measured signal *in vivo*. These include coil-specific receive profiles, spatial variations in the $B_0$ and $B_1^+$ fields, spatially nonuniform metabolite distributions, and realistic k-space encoding. While these spatial components are essential for generating advanced MRSI or multi-voxel synthetic datasets, they remain underrepresented across existing simulation tools.

### 3.2.1. Array Coils

Realistic simulation of array-coil acquisitions requires modeling spatially varying $B_1$- (receive) fields, coil sensitivity profiles, and inter-coil noise correlations. These factors determine how signal amplitude, phase, and noise vary across space and with different head positions inside the coil. Such variability is critical for replicating the acquisition characteristics of multichannel MRSI data. Although these parameters can be simulated by assigning coil sensitivity maps, coil loading factors, and noise correlation matrices, only two studies to date have explicitly incorporated these coil-specific effects into synthetic MRS data (Nelson, 2001; Rodgers & Robson, 2016). Wider adoption of array-coil modeling would enable more realistic benchmarking of reconstruction, denoising, and combination strategies.

### 3.2.2. Effects of $B_0$ and $B_1^+$ Inhomogeneities

Spatial inhomogeneities in the static magnetic field ($B_0$) and transmit field ($B_1$+) strongly affect MRSI signals. In earlier sections (2.2), line broadening, signal decay, and frequency shifts were discussed independently. However, spatially resolved modeling of $B_0$ inhomogeneities provides a unified and physically grounded way to simulate all three effects simultaneously.

$B_0$ field maps may be obtained from large MRI repositories, measured per scan, or measured once and fitted using smooth spatial basis functions such as polynomials. Several modern simulators implement $B_0$ inhomogeneity by over-discretizing the MRS voxel into many sub-voxels. For example, MRS-Sim simulates a three-dimensional voxel subdivided into a high-resolution grid and modulates each basis function with the complex exponential corresponding to the local $B_0$ offset, summing over the volume to produce the final line shape. This approach captures realistic intra-voxel dephasing beyond simple Gaussian broadening models.

Other tools, such as Osprey, do not explicitly model spatial $B_0$ distributions; instead, they optimize relaxation and line shape parameters that implicitly capture the effects of inhomogeneity. Additional simulation methods have been proposed for realistic $B_0$ modeling using high-resolution *in vivo* field maps or synthetic 3D fields (de Graaf et al., 2024; LaMaster et al., 2025; Lazen, 2020). Ideally, these approaches should also incorporate the true, non-ideal VOI selection profile and, in MRSI, the spatial response function (SRF, see Section 3.2.3) to achieve accurate spatial contamination patterns.

$B_1^+$ inhomogeneities, although less frequently modeled, can be incorporated similarly by scaling simulated signals according to transmit field variations across space, which is particularly relevant at high field strengths.

### 3.2.3. K-space Sampling and gradient imperfections

MRSI data acquisition uses spatial encoding: the specific k-space trajectory and the imperfections of gradient hardware determine the spatial response function. The SRF governs how signals originating from one location spread into adjacent or distant voxels, which is especially consequential when strong nuisance signals, for example extracranial lipids, contaminate the measurement.

To simulate realistic MRSI data, the high-resolution spatial metabolite distribution can be transformed into low-resolution k-space using simulated gradient trajectories that incorporate temporal delays, amplitude errors, and trajectory distortions. This produces an SRF that matches *in vivo* behavior more closely than idealized phase encoding.

Despite its importance, only two publications have simulated non-phase-encoded trajectories, such as spiral or rosette readouts, while nearly all others have relied solely on idealized phase encoding (Iqbal et al., 2017a; Kadota, 2021/2025). Incorporating realistic k-space paths, especially those affected by gradient nonlinearities or eddy-current-induced distortions, remains an important next step for synthetic MRSI development (de Graaf et al., 2024; Iqbal et al., 2019; Kasten et al., 2013; Lazen, 2020; Maguire et al., 2015; Nelson, 2001; X.-P. Zhu et al., 2003).

Taken together, current simulation tools vary widely in their ability to incorporate spatially dependent effects. Some support voxel over-discretization or field-map-based modeling, while others rely on simplified parameter adjustments that do not capture true spatial heterogeneity.

To help guide tool selection and future development, here are some of the existing simulators, detailing which spatial components they implement and identifying key challenges, including coil modeling, realistic $B_0/B_1^+$ fields, and non-ideal k-space trajectories (AlvBrayan, 2023/2024; bernhard-strasser, 2020/2024; *MRSI Data Processing Challenge 2024*, n.d.; MRSI-HFMR-Group-Vienna, 2024/2025).

## 3.3. Sample-Specific Components

Sample-specific components capture biological and experimental characteristics that differ across species, developmental stages, tissue types, and pathological conditions. Incorporating these elements enables synthetic MRS data to support a wide range of experimental, clinical, and methodological applications. These factors become particularly important for the modality-specific use cases discussed in Section 4.1 Functional MRS (fMRS), Section 4.2 Diffusion MRS (dMRS), and Section 4.3 Magnetic Resonance Spectroscopic Imaging (MRSI), as well as the application contexts described in Section 5.1 Clinical Applications and Section 5.2 Preclinical Applications. Although many of these components are well characterized *in vivo*, only a limited number of synthetic data generators currently incorporate them.

Human and non-human species differ in linewidths, chemical shifts, macromolecular content, and metabolite concentrations, particularly at high field strengths (Genovese & Marjańska, 2024). Gray matter, white matter, and other tissue types exhibit distinct relaxation properties and susceptibility profiles that influence spectral appearance. Incorporating region-specific relaxation or partial-volume effects improves realism, especially for spatially resolved approaches. Metabolite levels and relaxation times change across development, aging, and disease. Temperature, pH, susceptibility gradients, and preparation conditions alter resonance frequency, relaxation, and linewidth, particularly in preclinical or *ex vivo* studies. Spatial variations in $B_0$ and $B_1^+$ fields influence linewidth, frequency offsets, and signal scaling.

## 3.4. Temporal Dynamics

Temporal dynamics in MRS reflect any process that causes the spectrum to change over time, and synthetic data frameworks must capture these changes to reproduce realistic experimental conditions. Such temporal variation can arise from biological mechanisms that are of primary interest, from experimental manipulations designed to elicit a measurable response, or from unwanted instabilities originating in the scanner, the subject, or the acquisition itself. Importantly, these processes unfold across different timescales, and the dominant timescale observed depends on both the physiology under study and the experimental paradigm applied. This distinction is relevant for within-session paradigms such as functional MRS (fMRS), isotopic labeling studies, hyperpolarized experiments, magnetization/saturation transfer investigations, and diffusion-weighted acquisitions, as well as for between-session, or longitudinal, measurements in which inter-session variance plays a different and often larger role (Apšvalka et al., 2015; Fichtner et al., 2017; Lanz et al., 2013; Mullins, 2018; Bednařík et al., 2015; Mangia et al., 2007; Mangia et al., 2007a).

Biological or metabolically driven temporal changes are those that synthetic datasets are typically designed to emulate. In fMRS, for example, neural activation can lead to task-related changes in metabolite levels, accompanied by small BOLD-related linewidth or frequency modulations (Section 4.1). Isotopic labeling studies similarly require modeling of concentration trajectories governed by chemical exchange and metabolic flux, while hyperpolarized MRS produces rapid, large-amplitude decays dictated by the short $T_1$ relaxation times of the hyperpolarized substrate. These mechanisms generate interpretable temporal structure and represent the “desired” component of temporal dynamics.

A second category of temporal variation arises from the experimental design rather than biology. Block-design and event-related fMRS paradigms, inversion-recovery sequences for $T_1$ mapping, or diffusion-weighted MRS protocols with systematically varied gradient strengths (Section 4.2) all impose predictable temporal patterns on the acquired data. Multichannel acquisitions may further introduce slow changes in relative coil phase or SNR due to coil loading differences or subtle shifts in head position. Dynamic MRSI protocols extend these effects spatially, producing metabolite maps whose temporal evolution reflects both physiology and the specifics of the acquisition strategy. These elements represent deliberate, experimenter-controlled variability that synthetic data must represent when used to test analysis methods under realistic conditions.

In contrast, several forms of temporal variation are unwanted and act as noise sources that obscure the biological effects of interest. Scanner-related instabilities include both rapid, within-echo fluctuations and slow, session-level drifts. Almost all instabilities are predominantly driven by eddy currents generated by rapidly switching gradients, producing phase evolution or lineshape distortions that can be approximated using one- or two-parameter models (Craven et al., 2024; Marjańska et al., 2022; Simpson et al., 2017) or through irregular lineshape convolutions (Oeltzschner et al., 2020; Zöllner et al., 2024). Slow timescale changes arise from gradual heating of gradient coils and passive B0 shim components, leading to magnetic field drift across the experiment. These effects can be simulated by sampling frequency offsets across successive transients, as implemented in MRS-Sim (LaMaster et al., 2025). Physiological fluctuations further modulate phase, frequency, and amplitude on timescales corresponding to respiration, cardiac pulsation, or motion, and become increasingly prominent at high magnetic fields (Alves et al., 2024) or with specifically susceptible pulse sequences (Felblinger et al., 1998). Incorporating these instabilities into synthetic datasets enables systematic testing of correction strategies and is essential for evaluating the robustness of preprocessing and quantification pipelines.

Real experiments often exhibit combinations of the above mechanisms, and several recent approaches therefore emphasize multidimensional modeling. Rather than simulating the time dimension independently across spectral components, these methods jointly model chemical kinetics, relaxation processes, and temporal covariance across spectra, allowing time-resolved fitting algorithms to exploit structured relationships in the data (Clarke et al., 2024). As synthetic MRS frameworks continue to evolve, the ability to reproduce these interacting processes, both the desired biological effects and the accompanying unwanted instabilities, will be critical for generating datasets. These will serve as realistic, discriminative test sets for emerging analytical methods.

## 3.5. Challenges and Next Steps

Despite considerable progress in synthetic MRS data generation, several challenges remain, particularly when attempting to capture domain-specific nuances with sufficient accuracy. One unresolved issue concerns the scaling and characterization of macromolecular and lipid signals, which vary across brain regions, field strengths, and acquisition methods, yet remain insufficiently standardized for simulation purposes. Similarly, dynamic responses such as functional changes, diffusion sensitivity, and metabolic turnover are only partially represented in current frameworks, even though these effects are essential for the modality-specific applications described in Section 4.1 Functional MRS, Section 4.2 Diffusion MRS, and Section 4.3 MRSI.

A second set of challenges relates to spatial heterogeneity. Spatial variation in metabolite levels, relaxation parameters, and macromolecular content is still not fully understood, limiting the ability of synthetic data generators to accurately model regional specificity. While vendor differences and sequence-specific features can be accounted for in basis set simulation and signal modeling (Bell et al., 2022; Hancu, 2009), reproducing voxel-level variation requires incorporating $B_0$ and $B_1$ fields, coil sensitivity profiles, and partial-volume effects. These components are particularly important for the 2D and 3D MRSI applications discussed in Section 4.3, where realistic spatial contamination patterns and metabolite distributions are critical.

Multi-spectrum data generation represents another frontier. Many emerging techniques, including functional MRS, diffusion-weighted MRS, metabolic kinetics, relaxometry, and spectral fingerprinting, require large sets of related transients with controlled variability. Advances in computing resources, algorithmic efficiency, and understanding of *in vivo* variability will be necessary to support the creation of such comprehensive, multi-transient datasets.

Several physical effects also remain insufficiently characterized for reliable simulation. Baseline contributions result from a combination of macromolecular, lipid, and sequence-dependent effects and are not yet fully understood. Metabolite distributions are typically known only at coarse spatial resolution, which complicates efforts to generate realistic high-resolution MRSI datasets. Simulating difficult-to-quantify metabolites such as GABA or lactate adds another layer of uncertainty. Furthermore, the creation of realistic MRSI data often requires large intermediate data structures, posing computational and storage challenges for practical simulation workflows.

Future efforts should prioritize the development and sharing of standardized synthetic datasets and include underrepresented modalities such as hyperpolarized MRS or X-nuclei spectroscopy. Improving reproducibility across studies will require harmonizing simulation parameters, reporting standards, and methodological choices. The next major step involves integrating the diverse effects described across Sections 3.1 to 3.4 into unified simulation frameworks that incorporate scanner imperfections, frequency drifts, spatial heterogeneity, and temporal dynamics within a single model. Hard-to-characterize components such as baseline contributions and macromolecular variability also require further investigation to enhance realism.

Ultimately, progress will depend on broad collaboration across research domains, improved understanding of metabolic spatial and temporal dynamics, and the integration of advanced computational and analytical tools. These developments will collectively enable more realistic synthetic data, improve benchmarking of analysis pipelines, and support the expanding clinical and preclinical applications described in Section 5.1 and Section 5.2.

# 4. Modalities in MRS

Synthetic data have become an essential resource across multiple MRS modalities, enabling optimization of acquisition strategies, validation and benchmarking of analysis pipelines, and the development of data-driven models, including AI-based approaches. This section examines how synthetic data are applied within three major contexts: functional MRS (fMRS), diffusion MRS (dMRS), and MR spectroscopic imaging (MRSI). For each modality, we summarize how synthetic datasets are generated, their advantages and limitations, and the roles they play in advancing methodological development. We also outline the key challenges and future directions specific to each application, identifying gaps in current simulation frameworks and opportunities for improvement.

## 4.1. Functional MRS (fMRS)

***Considerations***

- ***Dynamic Changes in Concentrations and MR Visibility:*** *Time-resolved simulations should account for metabolite fluctuations arising from metabolic flux and/or compartmental shifts. These processes may occur on timescales ranging from sub-seconds to minutes and should be modeled according to the intended experimental paradigm (e.g., functional MRS, metabolic tracing, or hyperpolarized studies). Changes in MR visibility should be distinguished from true concentration changes where relevant.*
- ***Dynamic Linewidth and Relaxation Changes:*** *Time-dependent alterations in linewidth and relaxation parameters should be incorporated where appropriate. For example, BOLD-related effects may induce transient linewidth narrowing, which should be reflected in realistic functional simulations.*
- ***Confounding effects due to long acquisition time:*** *For prolonged acquisitions, simulations should include potential confounding effects such as gradual concentration drift, lineshape variability, frequency shifts, and phase instability. These effects may arise from subject motion, physiological fluctuations, or scanner instability and should be modeled when robustness to such perturbations is being evaluated.*

Functional MRS (fMRS) measures changes in metabolite levels over time, i.e. the temporal dynamics of metabolites in response to functional tasks, pharmacological interventions, or changes in physiology (see reviews (Jelen et al., 2018; Koush et al., 2022; Stanley & Raz, 2018)). Although many fMRS experiments operate on short timescales within a single scanning session, the concept more broadly encompasses any paradigm in which metabolite dynamics are interpreted relative to an experimental condition or state change. This distinguishes fMRS from longitudinal studies, where inter-session differences and longer-term variance dominate, even though both approaches examine metabolite responses to interventions or physiology. While fMRS is most commonly applied using $^{1}$H MRS, for example, lactate and glutamate levels that change in response to pain and visual stimulation (Mangia et al., 2007; Mangia et al., 2007a; Mullins et al., 2005; Prichard et al., 1991), dynamic MRS(I) has also been used with other nuclei such as $^{31}$P, $^{2}$H, and $^{13}$C (Duguid et a., 2025, García et al., 2025, Hu et al., 2009, Ma et al., 2022, Paul et al., 2025). Although some principles carry over, additional complexities, like pH-related peak shifts ($^{31}$P) or polarization dynamics ($^{13}$C), come into play. We acknowledge these X-nuclei approaches but focus here on $^{1}$H MRS, as no publicly available synthetic datasets are currently available for the other nuclei.

Synthetic data is playing an increasingly valuable role in fMRS and can be used to: 1) test hypotheses about the origins of neurometabolite fluctuations and their response functions, 2) assess the accuracy of both traditional and novel fitting methods in quantifying dynamic changes, 3) examine the impact of nuisance parameters (e.g., BOLD-induced linewidth fluctuations) on these measurements, and 4) explore the limits of temporal and spatial resolution in $^1$H fMRS acquisitions.

The key distinction between synthetic functional MRS and static MRS data lies in modeling neurometabolite fluctuations over extended acquisition times, along with associated confounding factors. While fMRS may be used to evaluate dynamics across longer time-scales (days) by comparing metabolite levels across multiple acquisition sessions, we focus here on the more typical scenario involving a time course of conventional or edited spectra in which one or multiple metabolites exhibit concentration changes within the scan duration. Key considerations in such simulations include the magnitude of the response (e.g., percentage concentration change) and its temporal characteristics (e.g., onset, peak, and decay profile). Commonly studied metabolites, such as lactate (Lac) and glutamate (Glu), have expected concentration changes in response to various functional stimuli that can be informed by findings summarized in recent meta-analyses (Mullins, 2018; Pasanta et al., 2023). While some studies suggest both shifts in metabolic flux and compartmental redistribution alter MR visibility, the exact mechanisms driving these changes remain unclear and controversial. Importantly, metabolite dynamics appear to reflect processes operating on different timescales, and the timescale observed depends on the experimental paradigm. For example, rapid glutamate changes have been reported to peak within approximately 0.5 seconds (Apšvalka et al., 2015; Mullins, 2018), whereas slower, sustained changes may reach a maximum around 120 seconds (Bednařík et al., 2015; Mangia et al., 2007).

An important consideration in synthetic fMRS is the inclusion of task-related linewidth changes resulting from BOLD-induced alterations in $T_2$*. While BOLD effects do not directly affect metabolite concentrations, they can influence spectral lineshapes in ways that can be misinterpreted as changes in concentration (Near et al., 2013) (Figure 3c - tNAA changes only due to linewidth changes). Several early fMRS studies have reported indicative ranges for these linewidth fluctuations (e.g., Bednařík et al., 2015; Mangia et al., 2007; Mangia et al., 2007a; Schaller et al., 2014), although their exact characteristics remain poorly defined (Schrantee et al., 2020). Motion presents another source of variability, introducing shifts in tissue composition within the voxel, changes in $B_1^+$ field homogeneity, and fluctuations in overall signal intensity over time. Additionally, scanner-related instabilities may cause gradual frequency drifts and imperfect water or lipid suppression that can accumulate over longer acquisitions. These gradual signal changes are particularly problematic for long block-design (>1 min) fMRS, where slow glutamate and lactate changes associated with prolonged stimuli can be difficult to distinguish from scanner drift and progressive head motion.

Advances in fMRS analysis methods have increasingly relied on synthetic data to evaluate and refine their performance under controlled conditions. Traditionally, fMRS studies grouped transients by block or event type and applied conventional spectral fitting methods, such as LCModel, to each group independently. However, newer approaches, such as 2D fitting techniques (Chong et al., 2011, Clarke et al., 2024; Kreis et al., 2005, Simicic et al., 2024; Tal, 2023) and direct linear modeling in the temporal domain (Morelli et al., 2024) (Figure 3d), offer improved precision by jointly modeling spectral and temporal dimensions. These methods help disentangle true metabolite responses from nuisance factors like baseline shifts, linewidth changes, or motion-related artifacts. Simulation studies have shown that 2D fitting can substantially increase precision, particularly in the presence of overlapping peaks, while also reducing the uncertainty of estimated response amplitudes (Clarke et al., 2024; Tal, 2023). These gains arise from jointly fitting multiple noisy spectra within a unified model, allowing shared parameters and explicit modeling of metabolite dynamics across time. This reduces noise-driven variance but may introduce bias if the assumed temporal model is violated. Similarly, regularization of fitting parameters, while proven effective in static MRS, has yet to be systematically explored in the context of fMRS (Wilson, 2025a). Synthetic datasets provide a powerful platform to test such strategies and define optimal analysis pipelines tailored to dynamic MRS data.

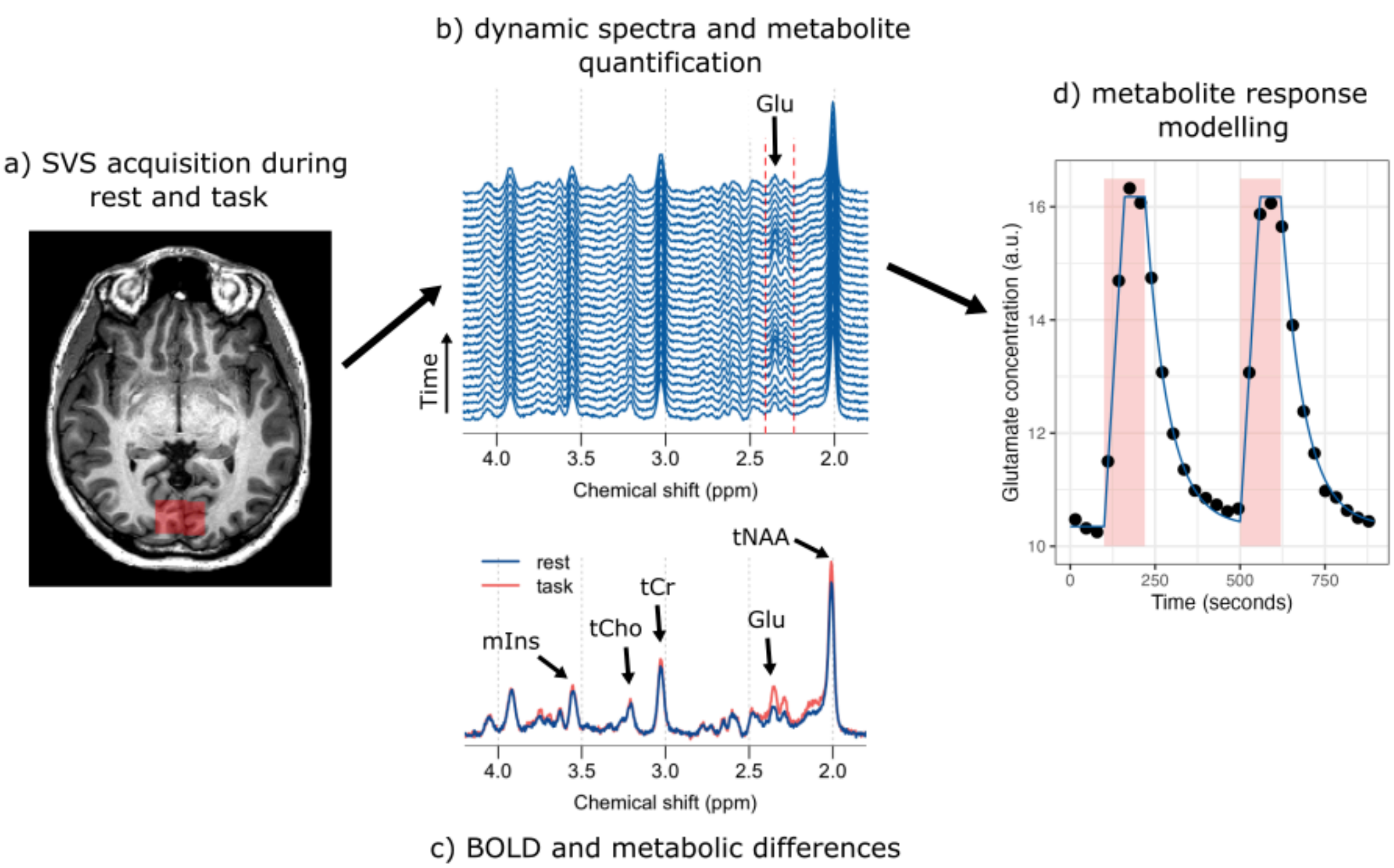

**Figure 3.** An overview of a typical fMRS experiment. **a)** Dynamic spectra are acquired from a single brain location whilst the participant performs a task. **b)** Spectra are analyzed to measure the concentration or percentage change in the target metabolites. **c)** Coincident changes in spectral linewidth from BOLD (or other artifacts) produce changes in spectral intensity that may be confused with changes in metabolite concentration. In this simulated example, the changes in glutamate are real, whereas the apparent change in tNAA (and other singlets) is purely due to a predicted change in linewidth from the BOLD effect. **d)** Analysis may be performed to temporally characterize metabolite responses or incorporate known or anticipated response functions to improve accuracy.

## 4.2. Diffusion MRS (dMRS)

***Considerations***

- *When generating synthetic diffusion-resolved MRS (dMRS) spectra,* ***artifacts specific to diffusion*** *acquisitions should be incorporated. These include phase and frequency drifts, signal intensity outliers, and other instabilities expected under in vivo diffusion-weighted conditions.*
- *To evaluate a processing pipeline and/or optimize an acquisition protocol, quantum-mechanical (QM) simulation of both the spectral and* ***diffusion dimensions*** *should be performed using the specific MRS sequence and an appropriate forward analytical signal model.*
- *When the objective is to assess sensitivity of the acquisition framework to a specific biophysical parameter (e.g., soma radius), generation of full synthetic dMRS spectra may not be necessary. In such cases, Monte Carlo simulations of metabolite diffusion within synthetic cellular geometries may be sufficient and more computationally efficient.*

$^1$H diffusion-weighted MRS (dMRS) uniquely combines the properties of $^1$H MRS (spectral/chemical shift) and diffusion-weighting acquisitions (b-value, diffusion time) to measure the diffusion properties of mostly intracellular metabolites and, as such, inform on the cell-type specific microstructure. For an overview of dMRS, refer to the consensus and recommendations paper (Ligneul, Najac et al., 2024) and dMRS reviews (Palombo et al., 2018; Ronen & Valette, 2015), as well as resources on MRSHub (*https://forum.mrshub.org/c/study-experiment-design/everything-about-diffusion-weighted-mrs/36*). dMRS simulations are pivotal to further our understanding of the complex relationship between tissue microstructure at the cellular scale and the encoded dMRS signal. In addition to the chemical shift and quantum interactions specific to MRS (with related artifacts and challenges), stochastic dynamics (e.g. diffusion and relaxation) encoded with specific RF and gradient schemes should be included in simulation frameworks.

To the best of our knowledge, Table S1 compiles the references using synthetic dMRS data and available repositories with open-source simulation materials. These references can be separated into 2 categories: **i)** the ones in which an analytical biophysical diffusion model is applied to each metabolite signal simulated with a (d)MRS sequence (Adalid et al., 2017; Christopher W Jenkins, 2021; Clarke et al., 2024; Mosso et al., 2022; Najac, et al., 2022) (Table S1), and **ii)** the ones creating numerical simulations of metabolites' molecular diffusion in substrates mimicking neural cells (Ligneul et al., 2024; Palombo et al., 2016, 2019)(Table S1). An overview of simulation strategies for dMRS data is shown in Figure 4.

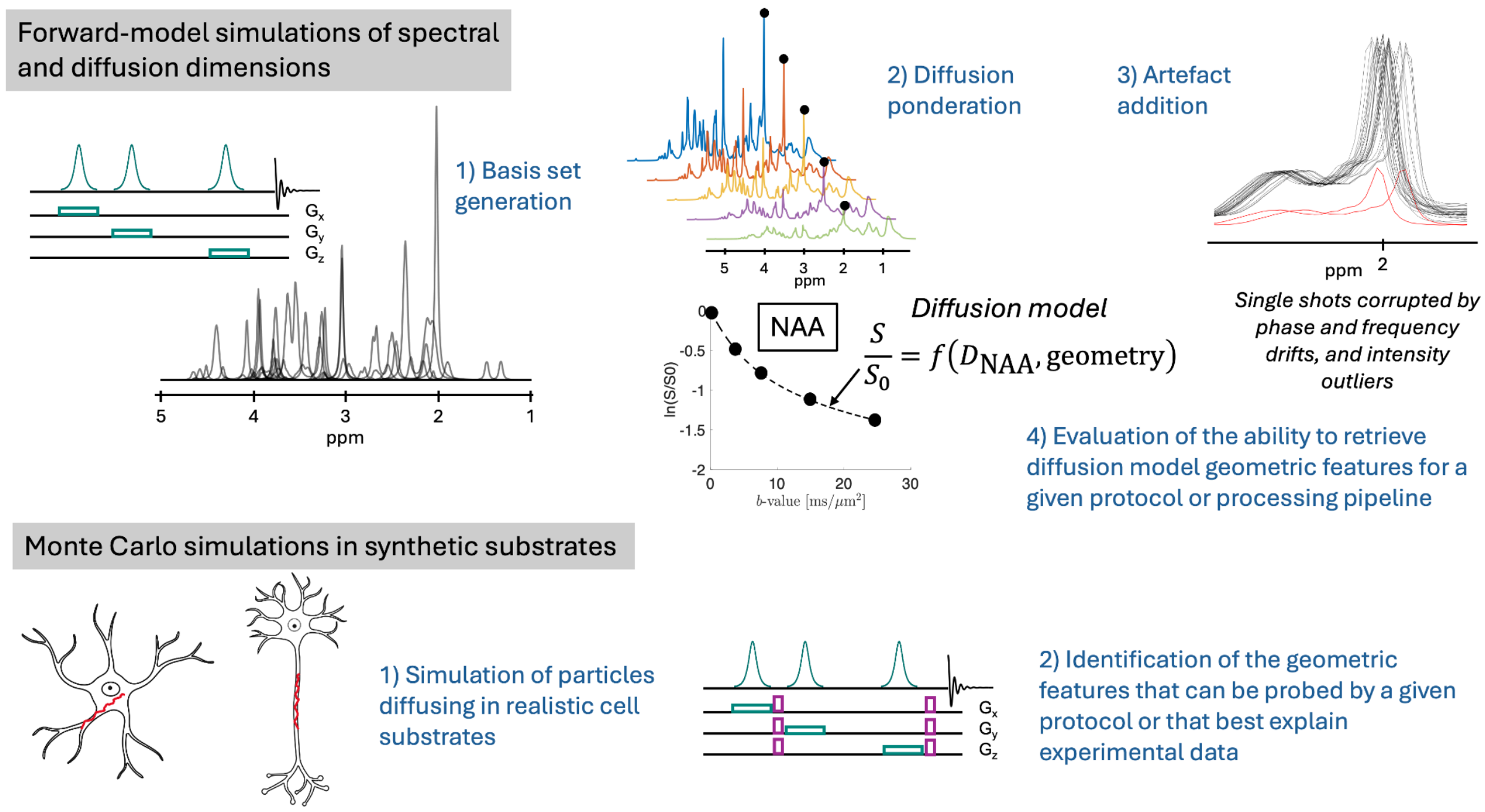

**Figure 4.** Overview of simulation strategies for diffusion-weighted MR spectroscopy (dMRS). **Top:** Forward-model simulations combine spectral and diffusion dimensions. **1)** A basis set is first generated for the chosen (d)MRS sequence, followed by **2)** diffusion weighting according to a geometric diffusion model (e.g., metabolite-specific signal attenuation as a function of b-value). **3)** Realistic artifacts, including frequency and phase drifts and intensity outliers, are then added to emulate single-shot experimental variability. **4)** This framework enables evaluation of how well a given acquisition protocol or processing pipeline can recover diffusion-related geometric features. **Bottom:** Monte Carlo simulations in synthetic cellular substrates, where **1)** particles diffuse within realistic neuron- or glia-like geometries, are used to identify which microstructural features are accessible with a **2)** given protocol and/or which best explain experimental dMRS data.

Pipelines from **i**) aim at evaluating protocols and processing/fitting approaches for typical *in vivo* dMRS acquisitions. Synthetic dMRS was used to highlight the benefits of 2D simultaneous fitting (of the spectral and diffusion dimensions) over 1D sequential fitting, the former increasing precision and accuracy of diffusion estimates and artifact corrections (Table S1: (Adalid et al., 2017; Clarke et al., 2024; Najac, et al., 2022), and compare phase/frequency shift correction methods (Table S5: (Christopher W Jenkins, 2021)). Additionally, synthetic data is key to developing and validating denoising algorithms, especially for low-SNR modalities such as dMRS. Principal component analysis (PCA) denoising was tested on simulated dMRS spectra and showed mixed results. While it improved the precision of diffusion estimates within a cohort, it also showed a deleterious homogenization of diffusion properties across metabolites, dependent on the inclusion of artifacts (Table S1: (Genovese & Marjańska, 2024; Mosso et al., 2022)).

Pipelines from **ii**) aim at evaluating modeling approaches, typically to understand which features of the complex cellular morphology can be estimated using the diffusion of intracellular metabolites. Synthetic data are often generated by simulating how metabolites diffuse within realistic cellular environments. A common approach is to create digital cellular structures that reflect morphometric features observed in microscopy data and then run Monte Carlo simulations of metabolite random walks within these structures using a specified dMRS protocol and a range of diffusivities. The resulting signal decays provide a controlled set of diffusion-sensitive metabolite-specific signal intensities. These synthetic datasets can then be used to train machine-learning models to learn the inverse mapping, enabling the estimation of microstructural features, such as branching order, projection length, or soma size, from the measured dMRS amplitude decays. (Table S1: (Ligneul et al., 2024; Palombo et al., 2016, 2019). These pipelines have so far been used without generating synthetic MR spectra.

## 4.3. MR Spectroscopic Imaging (MRSI)

***Considerations***

- ***Spatial variation*** *of metabolite concentrations should be anatomically meaningful and consistent with reported literature values for the tissue type and field strength under investigation.*
- *Field-strength–dependent spatial variations in magnetic fields (**$B_0$ and $B_1$**) should be modeled to reflect their influence on spectral appearance, including linewidth, frequency offsets, and signal amplitude heterogeneity.*
- ***Lipid and residual water*** *artifacts should be simulated with realistic spatial propagation characteristics. In particular, the spatial response function (sometimes referred to as point spread function, PSF) of the acquisition should be taken into account to ensure that contamination patterns reflect in vivo behavior rather than voxel-wise independent additions.*

Robust, high-quality MRSI processing methods that enable accurate and repeatable mapping of metabolite levels or concentrations are essential for clinical applications. To identify the best among the many proposed approaches, they must be compared to identical datasets with known ground truth. Openly available synthetic data provide a practical solution to this dilemma and would accelerate scientific exchange and the development of MRSI processing methods.

Most physical effects present in MRSI are also inherent to single-voxel MRS and are discussed in Section 2 Signal Models and Section 3 Advanced Components. These include chemical shift, J-coupling, $T_1$ and $T_2$ relaxation, $B_0$ inhomogeneities, $B_1^+$ and $B_1^–$ field variations (e.g., in array coils), residual water and lipid signals, and physiological or scanner-related instabilities. However, only a subset of these, primarily J-coupling and relaxation behavior, can be simulated in a manner similar to single-voxel MRS. Other effects must be modeled differently in MRSI because of the typically larger encoded volume, the spatial subdivision into many voxels, and the stronger spatial variability across the field of view.

In addition, MRSI requires explicit modeling of k-space sampling and spatial distributions, as described in Section 3.2 Spatial Components, which contrasts with single-voxel MRS that can be treated entirely in the time or frequency domain and has sharply defined spatial boundaries. While a single voxel can often be optimized for $B_0$ and $B_1$ field homogeneity before acquisition, even well-optimized MRSI acquisitions commonly exhibit substantial spatial variation in these fields across a slice or volume. These spatial dependencies must therefore be incorporated into synthetic data generation to achieve realistic simulations.

To date, the majority of publications on MRSI simulations have accounted for the spatial *in vivo-like* dependence of metabolite concentrations (Alves et al., 2024; Bjørkeli et al., 2025; Clarke & Chiew, 2022a; Iqbal et al., 2019; Lam et al., 2020; Maruyama et al., 2025; Motyka et al., 2019; X.-P. Zhu et al., 2003), but simpler assumptions have also been used (Haupt et al., 1996; Iqbal et al., 2017b; Kasten et al., 2013). In contrast, all aspects of realistic k-space encoding are typically neglected with the exception of exploring non-phase-encoded k-space trajectories (Iqbal et al., 2017b; Kadota, 2021/2025).

Regarding non-MRSI-specific effects, most publications incorporated spectral effects such as chemical shift, $T_2$ or $T_2$* decay, and J-coupling. However, some simulations were limited to singlet Lorentzian peaks. Seven publications modeled water nuisance signals (Alves et al., 2024; bernhard-strasser, 2020/2024; de Graaf et al., 2024; Kadota, 2021/2025; MRSI-HFMR-Group-Vienna, 2024/2025; Nelson, 2001; X.-P. Zhu et al., 2003), typically as a single Lorentzian singlet, without scaling to account for spatial variations in water-suppression efficiency. Some considered variable water suppression due to off-resonance and flip angle errors or simply by adding Lorentzian lines of variable amplitude but with fixed chemical shift. Lipid signals were simulated in only five publications (de Graaf et al., 2024; Kadota, 2021/2025; Lazen, 2020; Motyka et al., 2019; MRSI-HFMR-Group-Vienna, 2024/2025) .

Eleven studies considered $B_0$ inhomogeneities (Alves et al., 2024; de Graaf et al., 2024; Kasten et al., 2013; Lam et al., 2020; Lazen, 2020; Motyka et al., 2019; Nelson, 2001; X.-P. Zhu et al., 2003), but just two simulated $B_1^+$ inhomogeneities (de Graaf et al., 2024; Lazen, 2020) and another two addressed $B_1^-$ inhomogeneities (Nelson, 2001; Rodgers & Robson, 2010) analysis. Notably, only one publication so far has investigated the effects of subject motion (Motyka, et al., 2024).

The spatial nature of MRSI adds complexity that SVS does not need to address. These additional considerations make simulating MRSI datasets more challenging. Figure 5 highlights the spatial nature of MRSI and illustrates additional considerations that differentiate MRSI from SVS. The following aspects are fundamental in MRSI.

### 4.3.2. Prescans and Prior MRI Data

Many MRSI processing steps rely on prior knowledge obtained from MRI scans performed in the same session. Thus, it is important to also simulate prescan data (e.g., $T_1$w-MRI, $B_0$, $B_1^+$, coil sensitivities) with a typical range of errors along with the synthetic MRSI data. This point has been partially addressed (de Graaf et al., 2024; Nelson, 2001), but is sometimes replaced by using actual, measured *in vivo* data.

### 4.3.3. Motion and Breathing Artifacts / Physiological Instabilities

Voluntary motion and breathing are major sources of instabilities, which cause smearing of nuisance signals, even to distant voxels (Andronesi et al., 2021). This makes appropriate simulations of motion essential. High-temporal-resolution *in vivo* data, e.g., using optical tracking or EPI-navigator data, provide realistic motion patterns. Ideally, this also includes changes in $B_0$- and $B_1^+$-fields that are caused by participant motion. Only one abstract simulated this effect for MRSI (Motyka, et al., 2024).

## 4.3.4. Spatial Metabolic Distributions

For realistic MRSI data, it is essential to reflect the variability over a range of *in vivo* human metabolite concentrations and anatomical properties and account for population differences (e.g., age, sex, and pathologies). Most of the publications simulated spatial metabolic distributions (e.g., GM/WM/CSF contributions or virtual phantoms) in some form.

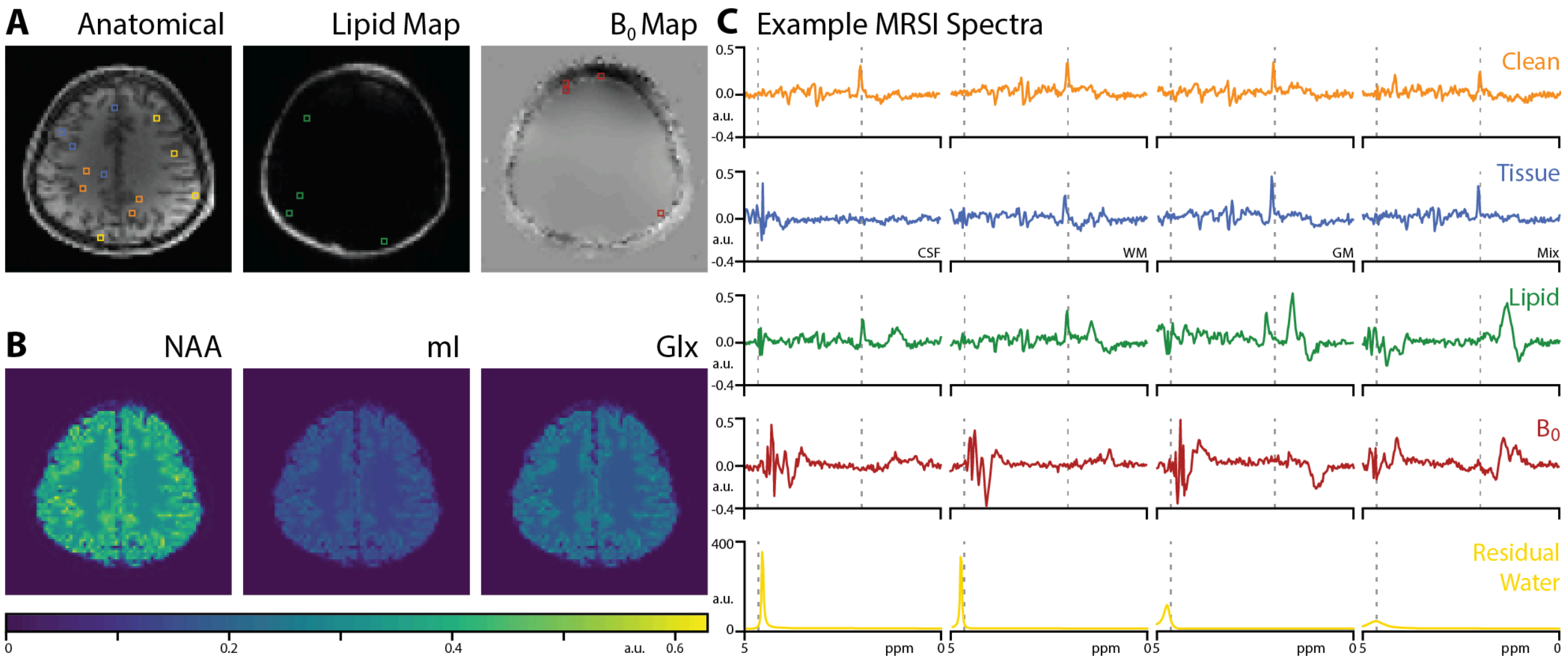

**Figure 5:** Illustration of MRSI data simulation using the MRSI Data Processing Challenge from the ISMRM Workshop on MR Spectroscopy, 2024. **A)** The anatomical, B0 map, and Lipid map. **B)** Metabolite maps for NAA, mI, and Glx. **C)** Example spectra from 20 voxels were produced by isolating the residual water using HSVD, isolating the lipid to down scale by a factor of 100x by removing residual water from the combined nuisance signal, and finally subtracting the nuisance from the full signal, correcting the phase, and combining with updated lipid signal. These spectra demonstrate clean (**Orange**), differing tissue types (**Blue**), variability in lipid artifacts (**Green**) and field homogeneity (**Red**), and the scale and shape of the residual water (**Yellow**). The spectra are ordered from left to right as they appear in the brain maps from top to bottom shown in A.

## 4.4. Challenges and Next Steps

### 4.4.1. fMRS

For functional MRS, three key areas of uncertainty were identified in the modeling of dynamic metabolite changes:

#### 4.4.1.1. The metabolite response function

While early findings provide some indications, the precise nature of dynamic responses, particularly for key metabolites such as Glu, remains speculative (Mullins, 2018). For many other metabolites, this response is even less well-defined. Since meaningful simulations of transient-to-transient metabolite fluctuations depend on an accurate response function, refining our understanding in this area is a critical topic of ongoing research. Meanwhile, any synthetic data modeling tools should remain adaptable to accommodate future advancements in response function characterization.

#### 4.4.1.2. BOLD-related effects on spectra

While the BOLD effect follows the well-characterized hemodynamic response from fMRI studies, typically peaking around 5 seconds (X. H. Zhu & Chen, 2001), its influence on metabolite spectra is more complex. The BOLD effect introduces changes in MRS signal characteristics, particularly metabolite lineshape, which in turn may impact the fitting process and, consequently, the derived metabolite concentration estimates. Many studies assume that these effects result in a simple, global Lorentzian broadening. However, a combination of Lorentzian and Gaussian broadening, along with potential higher-order asymmetries, may be more accurate (Uludağ et al., 2009). Additionally, hypothesized compartmental shifts and other physiological changes could lead to metabolite-specific lineshape alterations that are independent of the BOLD effect itself.

#### 4.4.1.3. Confounding factors

Both subject motion and scanner instabilities have the potential to introduce unwanted spectral variance and, therefore, errors in metabolite estimates. At best, these errors will mask the genuine metabolite dynamics, reducing the sensitivity of fMRS. At worst, these artifacts may be coincident with anticipated metabolite changes, resulting in spuriously elevated estimates of metabolite level changes. For example, a strong pain stimulus may introduce involuntary movement, and therefore coincident spectral variance, which may be mistaken for a metabolic change.

Given that lineshape changes induced by BOLD fluctuations can mimic spurious metabolite changes (X. H. Zhu & Chen, 2001), improved susceptibility contrast in high-field MRS necessitates a more detailed characterization of these effects. To improve the realism of synthetic fMRS datasets, future simulation frameworks should incorporate more flexibility in modeling lineshape variations, including the ability to selectively apply different broadening mechanisms to distinct spectral signal models.

Beyond these issues, other dynamic confounds, such as subject motion, $B_0$ drift, and lipid and water instability, are rarely accounted for in existing synthetic fMRS datasets. Incorporating these factors will enhance the realism of simulations and provide more reliable testing grounds for modeling approaches.

Finally, we emphasize the importance of sharing experimentally acquired fMRS datasets to refine synthetic data accuracy. The impact of BOLD and other physiological confounds is likely to depend on multiple factors, including field strength, behavioral paradigms, and acquisition parameters. Thus, pooling data across multiple studies may be necessary to generate sufficient evidence to support or refute specific models.

### 4.4.2. dMRS

The complexity and heterogeneity of dMRS simulations are highlighted by the references in Table S1. The inclusion of artifacts in synthetic dMRS data appears necessary for an accurate evaluation of processing pipelines (Table S1: (Christopher W Jenkins, 2021; Najac, et al., 2022)). Such artifacts should be made b-value dependent when this is expected *in vivo* (e.g., stronger phase drifts and more motion-corrupted shots at high b-values) (Genovese et al., 2021), and their subsequent influence on processing and fitting results should be evaluated.

Future applications of synthetic dMRS data include evaluating the interplay between diffusion and relaxation properties, optimizing thresholds for outlier removal, and fine tuning acquisition parameters. For example, selecting the optimal b-value range or SNR to maximize sensitivity to a specific model feature under experimental constraints such as fixed scan time, maximum gradient amplitude, slew rate, or peripheral nerve stimulation limits. It should be noted that although analytical models of diffusion are useful to test processing pipelines with synthetic data, degeneracies and often invalid assumptions, such as the short gradient pulse approximation, may bias the inference of cellular microstructure from dMRS signals. Numerical simulations of dMRS signals in neural cell substrates can also guide the design of sequences and inversion algorithms and help generate more reliable datasets to train machine learning models (Table S1: (Ligneul et al., 2024; Palombo et al., 2016, 2019)).

Overall, a more unified dMRS simulator (including analytical modeling and Monte Carlo simulation approaches) is needed because the synthetic data resources found in literature so far are mostly application-specific. Inspiration for such a unified simulator can be found in the MRI (MRzero (Loktyushin et al., 2021)) and diffusion MRI literature (DISIMPY (Kerkelä et al., 2020), MCMRSimulator.jl (*Michiel Cottaar / MCMRSimulator.Jl · GitLab*, 2025)) and should be adapted to address the challenges specific to dMRS.

### 4.4.3. MRSI

MRSI faces significant challenges in achieving robust and high-quality processing methods essential for clinical applications, as objective comparison across methods requires standardized data with known ground truths. Synthetic MRSI data offers a potential solution by enabling fair and reproducible benchmarking. It can also support the development of machine learning models, which can reduce the time and computational cost associated with processing large MRSI datasets, but only when extensive representative training and validation data is available.

As discussed in Section 4.3, several studies have incorporated specific physical effects, such as B0 and B1+ inhomogeneities, into synthetic MRSI data generation. However, these implementations remain incomplete. The majority of studies reviewed in Section 4.3 overlook complex physical effects such as gradient imperfections, array coil simulations, non-Cartesian encoding, subject motion, and frequency drifts. Moreover, the interdependence of effects like chemical shift, $B_0$-inhomogeneity, and RF pulses has largely been neglected.

To address these limitations, the next step should involve integrating the various effects considered in prior studies into a comprehensive simulation. This should include incorporating well-understood factors like system imperfections and the interdependence of multiple effects. So far, no publication has addressed scanner instabilities such as frequency drifts, eddy currents, or other gradient imperfections in simulations. Complex interdependencies have also been largely overlooked, including factors such as water- or lipid-suppression efficiencies influenced by $B_0$- and $B_1$-inhomogeneities through pulse profiles, the effects of water and lipid suppression on metabolites, or the interaction between $T_1$-weighting of metabolites and $B_1$-inhomogeneities, which results in varying $T_1$-weighting for different $B_1$ values.

Simulations should also account for impacts such as k-space trajectory, filtering, and point-spread functions. Further investigation is needed on contributions to spectral baselines along with MM and lipid signals. These components remain poorly understood and consequently, cannot be adequately simulated.
Another unresolved challenge is the need for high-resolution metabolic maps over the entire brain, especially for low-concentration metabolites that could serve as a foundation for realistic simulations. Initial efforts should focus on generating 2D MRSI data to provide more control over physics-based data creation. The optimization process should be incremental, beginning with these factors and later expanding to include spatially correlated $B_0$, $B_1$, and eddy current imperfections. These steps should facilitate the transition to generating 3D MRSI data.

In summary, synthetic MRSI data has supported the development of advanced spectroscopic imaging techniques and validated complex pipelines, for example: Simulating spatial distributions of metabolites and $B_0/B_1$ inhomogeneities has been critical for evaluating quantification algorithms (Zöllner et al., 2024). Incorporating gradient imperfections and physiological instabilities into MRSI synthetic datasets ensures realistic k-space reconstructions. Also, modelling coil-specific receive profiles and noise correlations have improved accuracy in array coil setups (Rodgers & Robson, 2016). Future work should focus on sharing standardized synthetic MRSI datasets.

# 5. Applications

## 5.1. Clinical Applications

***Considerations***

- *Metabolite concentrations, relaxation parameters, and metabolite ratios should be tailored to reflect* ***physiology- or pathology-specific*** *metabolic profiles relevant to the intended simulation scenario.*
- *When simulating data for specific pathologies, parameter inputs should be derived from* ***well-powered, representative*** *datasets to ensure that simulated alterations reflect realistic population variability rather than anecdotal or underpowered findings.*

In clinical settings, synthetic data has two primary use cases. The first is validating software packages, such as spectral fitting methods, which use *a priori* assumptions to constrain the problem space. In such cases, soft or hard constraints may be violated by different pathologies, causing the methods to fail. The second is supplementing pathological datasets to develop and validate machine learning and deep learning models. In both cases, the synthetic data must be representative of the targeted pathologies. So far, synthetic data has been employed to model pathological metabolite profiles associated with neurodegenerative diseases, psychiatric conditions, and tumors, with notable examples including:

- **AGNOSTIC Dataset:** Synthetic data representing over 20 clinical profiles, including neurodegenerative, psychiatric, and trauma-related conditions (Gudmundson et al., 2023).
- **Oncometabolites:** Tumor spectra were simulated by adding contributions from 2HG and Cystat, increasing contributions from tCho, mI, Lac, lip09, and lip20, while decreasing contributions from tNAA and Glu (Davies-Jenkins et al., 2026).
- **Idealized Setting:** Metabolite concentrations were sampled independently and uniformly between 0 and twice the reference average concentration for a normal, healthy brain but tCho was extended to [0mM, 5mM] where elevated values above twice the reference average concentration represent tumor tissues (Rizzo et al., 2023).
- **Tumor Modeling:** Machine learning-generated synthetic data for healthy tissue, low-grade, and high-grade brain tumors (Olliverre et al., 2018).
- **2016 ISMRM Fitting Challenge:** Synthetic tumor datasets were generated in addition to healthy datasets by adding varying amounts of lipids and acetate (Marjańska et al., 2022).

Despite these advancements, the use of synthetic data in clinical settings remains limited, particularly for rare diseases and complex clinical scenarios. For instance, biological variability in brain tumors complicates synthetic data modeling, as overlapping peaks between classes and varying necrotic lipid contributions, particularly in glioblastoma, can obscure clear differentiation (Tate et al., 1998). Additionally, strong heterogeneity is often observed within individual patients and between patients with the same pathology (Tate et al., 2006). Only a few studies have generated synthetic MRS data tailored to clinical scenarios (Dikaios, 2021; Gudmundson et al., 2023; Ortega-Martorell et al., 2014). Using data from the INTERPRET project, a synthetic dataset with 24 metabolites, including citrate and 2-HG, was developed to differentiate metastatic brain tumors from glioblastoma (Dikaios, 2021). However, key necrotic lipid contributions were omitted, and some simulated metabolites like bHB which are less relevant in this context (Tate et al., 2006) were included. Another approach utilized the INTERPRET SV short TE dataset (20-32 ms) with Bayesian non-negative matrix factorization (NMF) to generate synthetic data (Ortega-Martorell et al., 2023). Here, one representative case from each of four brain tumor classes was expanded into 50 instances per class using two types of randomly generated mixing matrices (20% and 35% variability). This synthetic dataset was designed not to replicate the empirical population but to validate the ideal number of sources for Bayesian NMF extraction. Another approach was used in a doctoral dissertation, where a hill climbing algorithm was employed to construct a Bayesian belief network from a small Alzheimer's disease sample, and forward sampling was used to generate 1,000 synthetic examples, resulting in improved classification performance within that dataset (York & Terpstra, 2023). Finally, synthetic SV MRS data at short TE (30 ms) has been used for three-class brain tumor classification (healthy tissue, low-grade, and high-grade brain tumors) employing generative adversarial networks, deep convolutional GANs, and pairwise mixture models (Olliverre et al., 2018), using a highly class-imbalanced dataset of 137 SV MR spectra (120 in train set). Accuracy values were above 90% except for Deep Convolutional Generative Adversarial Network (DCGAN), although few details are provided about the numbers of generated spectra.

## 5.2. Preclinical Applications

***Considerations***

- ***1-H:***
  - ***$B_0$ and Gradient Strength:*** *$B_0$ field strength and gradient performance should reflect the high-field systems and strong gradients typically used in preclinical MRS acquisitions.*
  - ***SNR Scaling:*** *Signal-to-noise characteristics should account for the inherently lower SNR per transient associated with smaller voxel sizes commonly used in small-animal studies.*
  - ***Spectral Resolution and Linewidth:*** *Linewidth and spectral resolution parameters should be adjusted to reflect the higher spectral dispersion and field-dependent effects observed at preclinical field strengths.*
  - ***Macromolecules and Baseline:*** *Macromolecular contributions and baseline characteristics should be modeled according to preclinical acquisition conditions and field strength.*
- ***General:***
  - ***RF Pulse and TE Constraints:*** *Sequence simulations should account for short RF pulses and very short echo times (TEs) that are frequently used in preclinical implementations.*
  - ***X-nuclei Considerations:*** *Realistic signal behavior should be implemented for non-$^{1}$H nuclei commonly used in preclinical research, such as $^{13}$C, $^{2}$H and $^{31}$P, including appropriate relaxation properties and spectral characteristics.*
  - ***Noise and Motion Effects:*** *Physiological noise sources and motion-related distortions specific to small-animal experiments should be modeled where relevant to the simulation objective.*

Similar to human studies, proton and X-nuclei synthetic data have been generated and used in preclinical settings for research in mice, rats, and other animal models. However, preclinical MRS operates with higher $B_0$ field strengths, stronger gradients, shorter RF pulses, very short TEs, and smaller volumes of interest, among other differences from their human counterparts. As a result, acquired MR spectra will exhibit distinct SNRs, spectral resolutions, linewidths, and artifacts when compared to human data. Each of these features needs to be identified and considered in the preclinical context when using synthetic data.

While preclinical synthetic data must account for many differences from clinical data, it has been similarly used for tasks such as validation of new acquisition protocols, testing processing algorithms, and emerging quantification methods, all of which need comprehensive evaluations. For newly developed acquisition protocols, preclinical synthetic data has been used to assess signal loss due to J-modulation under different sequences and conditions (Deelchand et al., 2021; Deelchand, Henry, et al., 2015). Due to the small size of the rodent brain, preclinical $^{1}$H MRS suffers from reduced SNR, which, by design, becomes particularly problematic in lower-SNR techniques such as MRSI and dMRS creating a greater interest in synthetic data for testing denoising algorithms (Alves et al., 2024; Mosso et al., 2022). Nevertheless, recent advances in rodent MRS have demonstrated sub-microliter voxel localization at 11.7 T, allowing for reliable quantification of 19 distinct metabolites in the mouse cortex (Abaei et al., 2025). Such empirical data provide an invaluable benchmark for synthetic MRS frameworks that aim to capture the high-spatial-resolution conditions seen in preclinical neuroimaging. For model fitting, synthetic data has been used to compare acquired versus simulated basis sets (Cudalbu et al., 2008) as well as to assess the impact of using full versus parameterized modeling of experimentally acquired MM spectra and of the degree of baseline freedom on metabolite quantification (Cudalbu et al., 2007; Simicic et al., 2021). In addition to proton MRS, X-nuclei synthetic data has been more commonly used for preclinical work, including hyperpolarized MRSI (Hu et al., 2010) which also uses the underlying signal models more generally (Miller et al., 2018), novel reconstruction algorithms (Ohliger et al., 2013), advanced Proton-Observed Carbon-Edited pulse sequence design (Henry et al., 2006; Xin et al., 2010), and evaluating the necessity of proton decoupling for $^{13}C$ MRS (Deelchand et al., 2006).

Described briefly in the AI Section, preclinical synthetic data has also been used in the development of machine learning and deep learning approaches for the quantification of extracellular pH (Fok et al., 2022) and targeted quantification of overlapping metabolites (H. H. Lee & Kim, 2020). A more detailed description of these works can be found in Supplementary Table S1. Furthermore, synthetic data has been extensively used to validate acquisition and quantification methods in preclinical models such as:

- **High-Field Models:** Basis sets derived from 9.4 T rodent brain spectra have been used to simulate metabolic and relaxation dynamics (Marjańska et al., 2022).
- **X-nuclei Spectroscopy:** Simulations have validated techniques for tracking metabolic changes using [$^{13}C$]-labeled compounds (Lanz et al., 2013).
- **Physiological Modeling:** Synthetic datasets incorporating motion artifacts and field instabilities have improved the robustness of preclinical imaging pipelines (Alves et al., 2024).

Despite its success so far, preclinical synthetic data still lacks standardization and comprehensive models for species-specific nuances and experimental paradigms. Lanz et al., 2021, provides a detailed overview and expert recommendations regarding rodent models, the effects of anesthesia, and data acquisition protocols that should inform the simulations of synthetic preclinical MRS data.

## 5.3. Software Validation Methods

*Considerations:*

- *Synthetic datasets should be recognized as essential for definitive validation of MRS methodologies, as they provide a* ***known ground truth*** *and eliminate uncontrolled biases and noise sources inherent to experimental data.*
- *The design and characteristics of synthetic validation datasets should be tailored to the specific application under investigation. To ensure* ***reproducibility and comparability across studies****, transparent data sharing and adherence to reporting standards are strongly recommended.*
- ***Large, in vivo–like synthetic datasets*** *should be developed as accessible community resources for benchmarking emerging methods. Such datasets should be derived from generative models of sufficient biophysical and acquisition complexity to ensure realistic variability. The development of a widely adopted reference synthetic dataset for benchmarking new software would represent a valuable addition to the existing MRS methodological infrastructure.*

The field of MRS is rich with methodological development, particularly in the arenas of data processing and modeling. While the end goal of this research is application to *in vivo* spectra, researchers cannot fully evaluate the accuracy and precision of their new methods without a known ground truth with which to compare. Synthetic spectra provide such a ground truth, allowing researchers to uncover potential biases inadvertently introduced by their methods.

As a result, synthetic data have been deployed in a diverse array of applications for validating new software tools. New data processing methods often leverage this approach to validate their accuracy, for example, in correcting shot-to-shot frequency and phase drift (Near et al., 2015; Simicic et al., 2024; Wilson, 2019), residual water removal (L. Lin et al., 2019), and denoising (Chong et al., 2011; Wilson, 2021), to name just a few. Likewise, the complex interplay of individual linear combination modeling (LCM) components necessitates stringent testing of new methods. These have included the optimization of the baseline definition and basis set composition (Chong et al., 2011), optimal model parameter regularization (Wilson, 2025b), and even the establishment of entirely new LCMs (Borbath et al., 2021; Clarke et al., 2024; Oeltzschner et al., 2020; Wilson et al., 2011; Zöllner et al., 2024). Indeed, the latter is a particular concern for the MRS community, with different LCMs producing distinct metabolite amplitude estimates (Zöllner et al., 2021). Synthetic data have also been used to validate more specific modeling assumptions or conceptual approaches, such as the separation of choline-containing compounds (PC and GPC) from total choline signals (Lindner et al., 2017).

The need for validation and benchmarking has prompted community-driven efforts such as the MRS fitting challenges (Marjańska et al., 2022; *Program — ISMRM 2024 Workshop on MR Spectroscopy*, n.d.). While these and other validation datasets cover spectra with various features, the datasets themselves are limited in size (Anderson et al., 2009). This restricts the conclusions that can be drawn from them as benchmarking datasets, particularly for data-hungry AI applications (Section 5.3). The scale and sheer diversity of the validation data needed might be met by other methods (Gudmundson et al., 2023, 2023a; Rizzo et al., 2023; van de Sande et al., n.d.), but it is crucial that the specific approach taken by researchers must be underscored by adequate reporting standards (Section 6.3) and, ideally, openly accessible data in standardized data formats (6.4 Data Format).

## 5.4 Acquisition Scheme Optimization

*Considerations:*

- *For [1]H MRS of the brain, include physiologically realistic macromolecular resonance signals in all simulations, with their own $T_1$ and $T_2$ relaxation dependence when possible, so that MM-metabolite interactions are captured as a function of acquisition parameters rather than assumed to be constant.*
- *Optimize across a range of in vivo relaxation parameter values reflecting known biological variability across brain regions, age, and disease states, rather than at a single assumed point, and favor acquisition schemes whose performance is stable across that range.*
- *Model realistic in vivo lineshapes and a distribution of linewidths that reflect residual B0 inhomogeneity after shimming, particularly when the optimization is sensitive to the degree of spectral overlap between metabolites.*
- *Incorporate actual RF pulse shapes and spatial excitation profiles into simulations, accounting for chemical shift displacement error and $B_1$ inhomogeneity, rather than assuming ideal hard pulses or perfectly uniform flip angles across the voxel.*
- *If using CRLBs to define optimal acquisition parameters, beware that they may not be completely valid at the lowest SNR levels.*

Synthetic spectra serve as an indispensable tool for the design and optimization of MRS acquisition schemes, offering a controlled and reproducible basis for exploring parameter spaces that would be prohibitively expensive, time-consuming, or logistically impractical to survey through experimental measurements alone. The general strategy involves synthesizing a large ensemble of spectra across a wide range of acquisition parameters, most commonly varying TEs, but also linewidth, repetition time, and sequence-specific variables, and then applying quantitative criteria to identify the optimal parameter combination that best satisfies/achieves the experimental objective. That objective may be the selective enhancement of a target metabolite or set of metabolites, the suppression of interfering resonances, or the minimization of statistical uncertainty in concentration or relaxation estimates, typically quantified by the CRLBs.

For single-acquisition methods targeting a specific metabolite, the use of synthetic spectra to systematically map the dependence of spectral appearance on sequence parameters has enabled principled optimization that would otherwise rely on intuition or coarse experimental sampling. The PRESS sequence has been optimized in terms of overall echo time and its partition into first and second echo periods with the explicit goal of selectively emphasizing the 2HG signal for non-invasive glioma characterization (Ganji et al., 2016). In a conceptually analogous approach, PRESS acquisition parameters were optimized for the detection of NAAG with explicit attention to minimizing contamination from the glutamate signal, which partially overlaps the NAAG resonance (An et al., 2014). Within the domain of J-difference editing, the GABA-editing experiment has been the subject of particularly rigorous simulation-based analysis. Kaiser et al. used synthetic spectra to develop an understanding of spatial interference effects in PRESS-localized editing spectroscopy and subsequently conducted a detailed theoretical and experimental analysis of localized J-difference GABA editing at 4 T, examining the four-compartment spatial response and the influence of sequence timing on artifact structure, ultimately informing the design of extended acquisition schemes (Kaiser et al., 2007; Kaiser et al., 2008). The glutathione (GSH) J-difference editing experiment has been similarly optimized through simulation: Chan et al. employed spatially resolved density-matrix calculations to characterize the TE dependence of GSH editing-on, editing-off, and difference signal over a wide range of echo times, determining that, after accounting for in vivo $T_2$ relaxation, the optimal TE for GSH detection at 3T is approximately 120 ms, a finding with direct practical implications for MEGA-PRESS implementations (Chan et al., 2017). More broadly, Landheer and Juchem extended this approach by constructing synthetic spectra for seventeen brain metabolites across six commonly used localization sequences, including MEGA-sLASER for GABA, and using CRLB minimization to provide systematic echo time recommendations spanning the range of standard clinical and research MRS experiments (Landheer and Juchem, 2020).

The utility of synthetic spectra becomes even greater when the optimization target is not a single acquisition but a set of spectra acquired under varying conditions and subsequently analyzed in combination, since experimental mapping of a high-dimensional parameter space would require prohibitive numbers of metabolite phantom or in vivo measurements. An et al. examined a J-suppression module inserted into a PRESS sequence and used synthetic spectra to identify the best combination of SVS acquisitions for the simultaneous determination of metabolite concentrations and $T_1$ and $T_2$ relaxation times (An et al., 2017). Within the domain of 2D J-PRESS experiments, two complementary approaches to multi-acquisition optimization have been explored. Hatay et al. used synthetic spectra to identify optimal combinations of echo times whose subsequent summation maximally enhanced the distinction of otherwise strongly overlapping metabolite resonances (Hatay et al., 2023). Monte Carlo simulations of synthetic spectra provided additional confirmation of their selected protocol alongside in vivo validation. In a complementary framework, Bolliger et al. addressed the problem of simultaneously estimating concentrations and transverse relaxation parameters from 2D J-PRESS data by computing CRLBs across a large number of hypothetical multi-TE acquisition sets with irregularly varying echo times to be evaluated with simultaneous modeling, using the resulting CRLB landscape to identify the experimental design that jointly minimizes parameter estimation uncertainty (Bolliger et al., 2013). Rizzo et al. adopted a related strategy for single-shot multi-echo spectroscopy, documenting the benefits of the approach using both Monte Carlo synthetic spectra and CRLB analysis, alongside in vivo results (Rizzo et al., 2023a). Wang et al. further advanced the estimation-theoretic framework by applying an optimization protocol based on synthetic spectra to identify sets of echo times that maximized the differentiation of metabolites from neurotransmitters and subsequently applied the same approach to optimize $T_2$ estimation in a multi-parametric MRSI context (Wang et al., 2024). When using CRLBs to define optimal acquisition setups, their limited validity in low SNR regimes (Landheer and Juchem, 2021) should be considered.

The extension of synthetic spectrum-based optimization to MRS fingerprinting represents a natural frontier. Existing fingerprinting work has largely been restricted to singlet resonances, for which single-spin Bloch simulations suffice and the explicit generation of J-coupled synthetic spectra is unnecessary (Kulpanovich et al., 2021). As fingerprinting methods are extended to encompass J-coupled spin systems, where signal evolution is sensitive to the interplay of sequence timing, RF pulse shapes, and coupling topology, the role of synthetic spectra in scheduling optimization will become essential rather than optional.

## 5.5. Artificial Intelligence (AI) Applications

***Considerations***

- *The operational domain of AI models trained on synthetic data should be explicitly defined and characterized to identify performance boundaries, failure modes, and limitations.*
- *Synthetic training datasets should model clinically relevant metabolite distributions, nuisance signals, and acquisition artifacts to promote robustness under real-world conditions.*
- *Synthetic datasets should reflect the heterogeneity of in vivo acquisitions, including variability in hardware platforms, pulse sequences, acquisition protocols, physiological states, and demographic characteristics.*
- *AI models trained on synthetic data should be validated on independent in vivo datasets using clearly defined performance metrics and uncertainty quantification to assess clinical translatability.*
- *Benchmark synthetic datasets and standardized reporting protocols should be developed and adopted to improve reproducibility, transparency, and comparability across AI-driven MRS studies..*

Data synthesis and augmentation have increased in popularity due to limited access to *in vivo* data. A common strategy is to randomly select a partition of the number of signal averages or spectral transients (Berto et al., 2024; Chen et al., 2023; Dias et al., 2024; H. Lee et al., 2020). Related bootstrap-based approaches have also been used, in which new spectra are constructed by randomly recombining existing spectra point-by-point, effectively resampling the empirical distribution to create multiple synthetic realizations from a limited number of acquisitions (e.g., Bolliger et al., 2013). While these techniques create new spectra, they do not increase variability in linewidth, noise, or metabolite concentrations and ignore properties of single-shot spectra such as variations from phase cycling schemes or spectra editing schemes. Therefore, it is a form of augmentation that does not address generalizability or public availability (Gudmundson et al., 2023). Other augmentation strategies include multiplying and then corrupting *in vivo* spectra with noise, phase and frequency realignment, and linewidth broadening (Shamaei et al., 2023). These approaches can bridge part of the gap and increase the robustness of AI models, but for broader applications, such as in clinical populations, they still do not diversify the data sufficiently and may risk overfitting to sparsely available *in vivo* data.

Alternatively, the use of purely synthetic data for training and testing AI methods for MRS applications is widely adopted (see Table S1). Simulated data offer several advantages, including the creation of large quantities of data, increased parameter diversity, and, critically, access to known ground truth values for accurate evaluation and benchmarking. Synthetic data also enhance accessibility due to their cost-effectiveness, allow the generation of disease-specific spectra, and enable class-balanced datasets.

As highlighted in Table S1, many past studies have employed simulation techniques, yet these often lack comprehensive modeling of background signals and nuisance artifacts. Many studies rely on basic simulations such as phase and frequency shifts, linewidth modulations, Gaussian noise, and simplified MM baselines. However, the inclusion of nuisance signals such as lipid contamination, residual water peaks, eddy-current effects, spurious echoes, and non-Gaussian noise is less common. Past works have largely focused on processing and quantification, with hardware-near approaches, realistic single-coil data, and individual transients with accurate correlations still lacking. Additionally, simulating nuisance signals such as water, fat, MM, and other baselines are intricate and therefore often omitted, hindering realistic reflection of *in vivo* conditions.

The importance of training data in AI is fundamental, as models rely solely on the data to learn meaningful tasks. Synthetic data therefore offer a way to generate diverse and extensive datasets for training. Recent literature shows a trend toward synthetic data, particularly for accessing ground truth values in quantification tasks where accuracy and precision metrics can be uniquely defined. However, these papers also highlight limitations associated with current simulation methods, especially in accurately capturing *in vivo* conditions. While polynomial baselines may make spectra look more realistic, they are not well suited for AI models, as neural networks can easily learn and remove such structures without improving robustness. Incorporating smooth random walks or more realistic baseline components can help alleviate this limitation. To ensure that a model performs well in real-world settings, the training distribution should encapsulate or surpass the testing distribution (Tate et al., 1998), minimizing distribution-driven bias. The working regime of an AI model, that is, the range of inputs over which it performs reliably, must be understood to identify limitations, optimize training, and improve reliability. Benchmark synthetic datasets and standardized reporting protocols will be essential for improving reproducibility, comparability, and the development of robust AI solutions in MRS.

## 5.6. Challenges and Next Steps

### 5.6.1. Clinical Applications

A key challenge in generating synthetic MRS data for clinical populations is the limited prior knowledge of biological variability specific to each condition. Unlike healthy populations, where variability patterns are better characterized, clinical datasets often lack comprehensive phenotyping, which complicates the creation of realistic and representative synthetic models.

One primary hurdle is incorporating additional basis set components to reflect a broader range of metabolites. While standard basis sets cover common metabolites, expanding them to include rare or condition-specific metabolites is crucial for simulating pathological conditions and rare clinical scenarios. For instance, extending metabolite ranges as depicted in studies by Lee & Kim (2018, 2022) can provide more robust validation frameworks. But as shown by Rizzo et al., 2023, caution remains necessary because machine learning and deep learning models may behave unpredictably when exposed to data distributions that differ from their training domain.

Beyond altered metabolite concentrations, structural pathology introduces additional complexity that is often not represented in synthetic datasets. Lesions, necrotic tissue, hemorrhage, and byproducts from surgical or other medical interventions can substantially alter water content, relaxation rates, and macromolecular background contributions. These factors may invalidate assumptions that are typically appropriate for healthy tissue and can introduce severe baseline distortions that exceed those commonly observed in normative data. If such effects are not modeled, synthetic data risk underestimating the spectral variability encountered in real clinical populations.

In certain pathological contexts, major metabolite peaks may be markedly reduced or effectively absent. This has direct implications for both simulation design and reporting conventions. For example, it may not be meaningful to report signal to noise ratio for NAA or linewidth for creatine if those metabolites are not present or fall below detectable levels in the simulated spectrum. Clinical synthetic data frameworks therefore require explicit consideration of scenarios in which canonical spectral features are altered, suppressed, or missing altogether.

Clinical variables substantially influence metabolite concentrations and relaxation values. As outlined in the preceding section on physiological metabolite ranges (Metabolite Ranges), factors such as brain region (Pouwels & Frahm, 2005) and tissue composition (Harris et al., 2015) systematically shape spectral properties and baseline concentration distributions. In clinical contexts, however, these sources of variability are not merely descriptive but directly affect diagnostic thresholds, group comparisons, and biomarker interpretation. Age (Angelie et al., 2001; Gong et al., 2022), sex, and other demographic factors alter metabolite baselines and relaxation times in ways that may overlap with disease effects. Similarly, physiological state, medication status, psychiatric or neurological conditions, and psychological factors introduce structured variation that can mimic or mask pathology.

Synthetic datasets that do not explicitly model this multidimensional variability risk encoding overly narrow healthy reference assumptions, thereby limiting their translational validity. For clinical applications such as algorithm benchmarking, artificial intelligence training, or biomarker validation, synthetic data generation should therefore incorporate stratified biological and demographic parameters that reflect real world patient heterogeneity. Doing so will improve external validity, reduce bias in downstream models, and enhance the clinical interpretability of methods developed using synthetic spectra.

### 5.6.2. Preclinical

In the preclinical field, several challenges remain regarding the generation and application of synthetic data. One major limitation is the lack of standardized frameworks for generating synthetic datasets tailored to species-specific characteristics and experimental paradigms. Unlike in human MRS, where spectral simulations and basis sets are generally well-characterized, preclinical studies must account for additional physiological and anatomical variations between species, such as differences in metabolite relaxation properties and $B_0$ field inhomogeneities. Establishing standardized methodologies for synthetic data generation would improve reproducibility and comparability across studies.

Another critical future direction is the integration of machine learning-based approaches for more sophisticated synthetic data generation. Emerging deep learning models have demonstrated the ability to create realistic MR spectra with varying noise characteristics and spectral distortions, providing a powerful tool for training robust quantification pipelines. These approaches could enhance preclinical MRS by generating more realistic datasets for training and validating AI-driven quantification methods, as has already been explored in applications such as extracellular pH estimation and overlapping metabolite quantification.

Additionally, there is a growing need to develop synthetic datasets that incorporate physiological variability, including motion artifacts, field drifts, and metabolic fluctuations over time. This would be particularly relevant for studies involving hyperpolarized MRS, where rapid metabolic changes necessitate high temporal resolution and precise modeling of kinetic parameters. Such advancements would enable synthetic data to better mimic *in vivo* conditions, leading to more accurate testing of acquisition and analysis methods.

Finally, collaboration between research groups and the establishment of open-access synthetic data repositories could accelerate progress in the field. Standardized synthetic datasets with well-characterized noise, artifact profiles, and metabolite dynamics would serve as valuable benchmarks for evaluating emerging acquisition protocols and quantification algorithms. By fostering a more unified approach to synthetic data generation and application, preclinical MRS research can continue to refine and expand its methodological toolkit, ultimately improving the reliability and translational value of its findings.

### 5.6.3 Software Validation

The challenges of using synthetic data to validate software can be broadly dichotomized into the challenge of data generation (i.e., the creation of data that reflects the variation and complexity of the in-vivo environment) and the challenge of accessibility and replication (i.e., the ability of subsequent studies to utilize similar methods of validation for comparison and consistency).

#### 5.6.3.1 Data generation

True validation of MRS methods using synthetic data is only valuable when the generated data sufficiently represent the intended application and the full range of expected variability. In other words, if the simulation is too simplistic or idealized, or relies on experimental parameters from a single study or site, the utility of the resulting validation may be diminished. Methods validated on such data might perform adequately within the confines of the study, but then break down when confronted with real in-vivo data. Bridging the gap between simulation and real data is one of the key challenges of using synthetic MRS data in this manner.

While it is difficult to lay out guidelines that apply across the broad spectrum of potential applications, researchers should carefully consider their choice of basis set simulation, signal model, parameter distributions, and representation of common artifacts. We recommend researchers include as much complexity in their validation data as they may reasonably expect from their application data, and should otherwise clearly report the limitations of their validation by reporting the full scope of parameters used (see Section 6.2). Furthermore, a robust sensitivity analysis is also important to determine the effects of variability or uncertainty of each parameter on the final synthetic data.

#### 5.6.3.2 Accessibility and replication

Another challenge is in the replication of previous benchmarking and validation studies. Synthetic data are often not openly shared alongside publications. As data generation is often done as a small part of a larger study, it can often be underreported in the methods. This limits the possibility of replication studies and comparability of future studies working on similar methods. This is most directly addressed with open science practices and reporting standards. An important future goal is the development of a standardized synthetic dataset that can be used to evaluate new tools, i.e., all new fitting algorithms could be tested on the same synthetic dataset that includes known standards (e.g., known concentrations) to evaluate accuracy or precision and enable direct comparison with existing software tools.

## 5.6.4 Acquisition Scheme Optimization

A central limitation in using synthetic spectra for acquisition scheme optimization is the degree to which simulation fidelity determines the translational validity of the resulting recommendations. Spin system parameters (chemical shifts, J-coupling constants, and relaxation times) are routinely drawn from a small number of reference datasets measured under controlled in vitro conditions, with Govindaraju et al. (2000) serving as the most widely adopted source. Subsequent work has shown that these values can deviate meaningfully from their in vivo counterparts (Near et al., 2013), yet sensitivity analyses examining how such uncertainties propagate into optimization conclusions remain the exception rather than the rule. Macromolecular and lipid contributions present a related problem: their $T_2$ relaxation times differ substantially from those of metabolites (Murali-Manohar et al., 2020), meaning their relative contribution to the observed spectrum changes as a function of the very acquisition parameters being optimized. Additionally, their variability across brain regions, age groups, and pathological states (Oz et al., 2021) is rarely reflected in simulation frameworks. Most acquisition optimizations are performed at a single assumed set of $T_1$ and $T_2$ values. Although regional and age-related differences in metabolite relaxation times within the healthy adult brain are generally modest (Mlynárik et al., 2001; Träber et al., 2004; Marjańska et al., 2012) and may not fundamentally alter optimal measurement parameters, field strength–dependent shifts (e.g., 3T vs 7T) are more substantial and can meaningfully influence the expected relaxation regime and thus the center of the optimization range.

RF pulse modeling represents another underaddressed source of discrepancy between simulated and observed performance. Most simulations assume ideal hard pulses and perfectly uniform excitation profiles, neglecting chemical shift displacement error, $B_1$ inhomogeneity, and slice profile imperfections. Chemical shift displacement error is particularly consequential at higher field strengths where spectral dispersion amplifies spatial offsets between coupled resonances (Scheenen et al., 2008), and for J-coupled spin systems, limited RF pulse bandwidth alters multiplet amplitudes and phases in ways that can only be captured through full density matrix simulations incorporating experimentally characterized pulses (Kaiser et al., 2008). CRLBs are asymptotically valid under correct model specification, but can become increasingly optimistic relative to actual fitting performance in low-SNR regimes or when model assumptions are violated, including in high-SNR conditions where subtle line-shape deviations or unmodeled effects dominate. If not carefully considered, this divergence can impact the sequence optimization. To mitigate this risk, independent empirical validation becomes essential. Monte Carlo validation at realistic noise levels can be an important complement to CRLB analysis and should be treated as a methodological standard rather than an optional addition, as demonstrated by several of the works reviewed above (Bolliger et al., 2013; Hatay et al., 2023; Rizzo et al., 2023).

Finally, the field lacks standardized frameworks and open benchmark datasets against which simulation-based optimization strategies can be systematically evaluated and compared. The current diversity of simulation tools, spin system databases, optimization criteria, and fitting pipelines makes cross-study comparison difficult and limits reproducibility. The development of shared, well-characterized synthetic datasets with documented noise properties, baseline models, and ground truth parameters would provide a common reference for benchmarking competing strategies, accelerating methodological progress and improving the reliability with which simulation-derived recommendations translate into in vivo practice.

### 5.6.3. AI in MRS

AI methods require extensive and well-characterized training and validation data. Because *in vivo* datasets are limited and lack true ground truth, synthetic data often enable AI development in MRS (van de Sande et al., 2023). A central challenge is that model performance depends on the quality of the synthetic training data and on how closely the training distribution matches the *in vivo* data encountered during deployment.

#### 5.6.3.1. Domain shift

Machine learning (ML) and deep learning (DL) models learn patterns in the training data under the assumption that these generalize to new inputs. When simulated data is oversimplified or fails to capture real-world variability, the training and testing distributions diverge, resulting in **domain shift** and degraded performance. This occurs, for example, when simulations lack realistic nuisance signals, spectral diversity, or acquisition-dependent variability (Gudmundson et al., 2023). Addressing domain shift requires improving the realism of synthetic data to better reflect *in vivo* scenarios, using ML architectures designed for stronger generalization, incorporating spectral priors for more robust feature learning, and detecting distribution shifts early in development (Tate et al., 1998).

Robust simulations are particularly important when training data is sparse, because AI performance is tightly coupled to how well the synthetic training distribution matches the target *in vivo* distribution. If the diversity of real data is not represented, models will fail to generalize. Hybrid models that combine data-driven and physics-informed components can enhance robustness, for example through DL-aided least squares optimization (Chen et al., 2024) that improves generalization to out-of-distribution metabolite levels. Incorporating differentiable spectral-parameter layers (Tu et al., 2025) can further improve robustness across scanner vendor numbers, signal averages, and spectral resolutions, as demonstrated in recent work on multivendor harmonization.

#### 5.6.3.2. Increasing Realistic Representation

Representing realistic single-coil data, transient-level correlations, and nuisance signals such as water, fat, and MM remains challenging. Many simulations still omit these components due to their complexity. Generative models such as GANs and VAEs (Rizzo et al., 2023) can help by learning the underlying distribution of real spectra and producing synthetic data that better mimic *in vivo* behavior. However, these models cannot provide data with known ground truth metabolite concentrations unless conditional generative control is available. Future efforts should prioritize more realistic simulation methods and careful integration of generative models. Relevant approaches in related fields, such as Reyes et al. (2020), may provide useful insights.

#### 5.6.3.3. Quality Metrics

Evaluating the quality of synthetic data is essential to ensure that it reflects real-world conditions and supports reliable AI model development. Beyond visual inspection, quantitative measures such as Kullback–Leibler divergence or other distributional metrics commonly used in generative modeling can be applied. Performance comparisons between models trained on synthetic versus real data also provide practical indicators of how well the synthetic data approximate *in vivo* conditions.

Relevant work in spectral similarity assessment, such as the method introduced by Mostafapour et al. (2026), offers additional tools for evaluating how closely synthetic spectra match real acquisitions. Such approaches can help quantify deviations in spectral shape, noise structure, or artifact profiles and may guide future improvements in synthetic data generation.

#### 5.6.3.4 Reproducibility

Several AI methods directly modify spectral data, for example, by denoising or artifact removal. It is essential to evaluate how such methods affect the reproducibility of biologically relevant outcomes, such as metabolite concentrations or ratios, rather than relying on improved visual appearance alone.

DL–based denoising and artifact-removal approaches require particular scrutiny. Ghosting-removal and denoising methods have been proposed (Dziadosz et al., 2023; Kyathanahally & Kreis, 2017), and while they can substantially improve the visual quality of spectra, their impact on quantitative accuracy must be carefully assessed. Similarly, methods designed to remove spurious echoes (Gudmundson et al., 2023, Song et al, 2023) or motion artifacts must demonstrate that they preserve or improve quantitative reproducibility in MRS and MRSI analyses. Ensuring that AI-based corrections enhance, not distort, the underlying biological signal is essential for their application in clinical research (Clarke & Chiew, 2022b; Dziadosz et al., 2023; Kyathanahally et al., 2018; H. H. Lee & Kim, 2020). Recent work by Merkofer et al. 2025 used a DL model to decompose spectra into their constituent components, including a nuisance component, which can then be assessed individually for their quality prior to complete reconstruction. This ensures the integrity of the components is maintained, instead of relying on global visual improvement of the spectra. Care must also be taken to consider potentially altered noise characteristics when DL methods modify spectra, possibly invalidating subsequent methods that rely on the assumption of uncorrelated, Gaussian noise.

#### 5.6.3.5. Additional Challenges and Next Steps

Remaining gaps include limited standardization of simulation parameters, insufficient realism of nuisance signals, and lack of shared benchmark datasets. Future developments should prioritize more diverse and realistic synthetic datasets, standardized reporting protocols, and community engagement. The accompanying master table of publications (Table S1 Master Table) is intended as a living, openly accessible resource, and contributions from researchers will help maintain its completeness and relevance.

# 6. Validation and Accessibility

## 6.1. Synthetic Data Validation Methods

***Considerations***

- *At least one representative synthetic dataset should be visually displayed when introduced in a publication to justify its structure and realism relative to the intended application.*
- *Visual inspection should be complemented by quantitative metrics that characterize similarity to in vivo or experimentally measured data. These may include **cross-correlation measures**, **signal-to-noise ratio (SNR)**, **full width at half maximum (FWHM)**, or other post hoc quality indices. Such metrics are particularly important in applications sensitive to bias, including the generation of training data for AI models.*
- *The selection of the empirical **reference dataset against which synthetic data are compared** should be carefully justified, as it may substantially influence perceived validity and generalizability.*
- *Future guidelines defining validation limits for synthetic spectra should remain adaptable to diverse and emerging applications. Both **qualitative and quantitative validation** approaches are associated with inherent biases, and no single validation strategy should be assumed to be universally sufficient.*

This section reviews current approaches used to assess whether synthetic datasets are appropriate for a given application in MRS. Here, the term “validation” is used to denote post hoc assessment of suitability for use. Features incorporated during dataset generation itself are not considered to constitute validation. For example, designing a synthetic spectrum to match the quality parameters and relative metabolite amplitudes of an in vivo measurement is not, in this context, a validation step. In contrast, quantitative comparison of the resulting synthetic data with the corresponding in vivo acquisition, such as by cross-correlation (Figure 6), represents post hoc validation.

To formulate this description, a non-systematic review of MRS literature was conducted of representative publications (Table S5) that used synthetic data to investigate a question within *in vivo* MRS. In addition, each member of the Validation Methods focus group reported their current synthetic data use case(s) and associated validation protocols (Table S6).

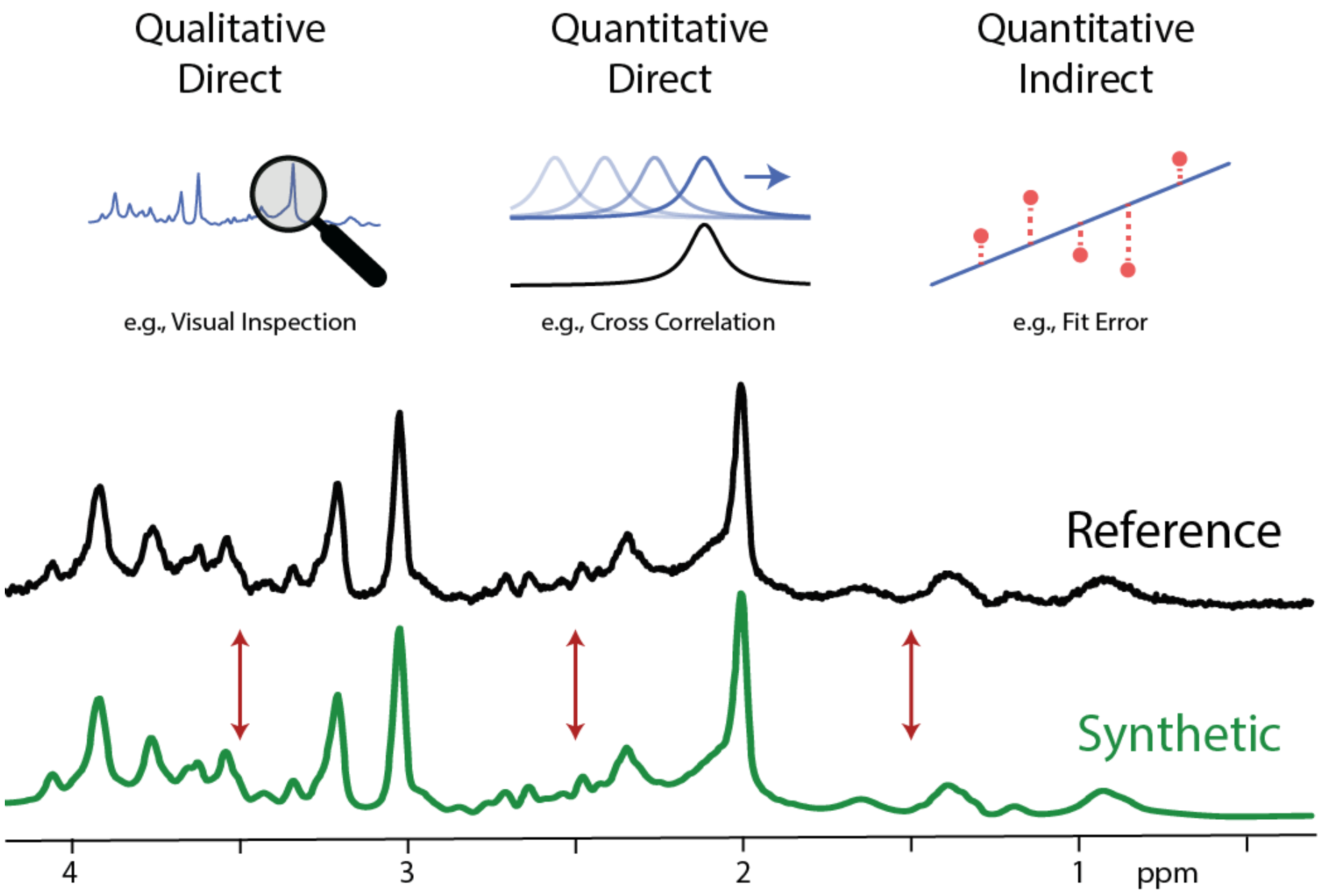

**Figure 6: Synthetic data validation framework.** A candidate synthetic dataset (green) is validated by comparison with a reference (black), which may consist of *in vivo* or phantom measurements or a conceptual target. Validation approaches include qualitative assessment (e.g., visual inspection), quantitative similarity metrics (e.g., cross-correlation or feature comparison), and indirect evaluation via downstream model performance on measured data (e.g. fit error).

The literature survey identified 39 publications documenting diverse applications of synthetic data. These included training ML models for various processing and analysis steps, providing ground-truth validation data for such models, providing ground-truth validation data for non-ML methods for processing and analysis (e.g., frequency registration, denoising, linear combination modeling), and providing context for making decisions about acquisition protocols (i.e., evaluating effects of variables like sequence timings or frequency drift on signals of interest). Similarly, this work considered 14 self-reported synthetic data application and validation protocols. Among them were applications for training, developing, and evaluating ML models; providing ground-truth metrics to evaluate diverse processing and quantification techniques, both ML-based and otherwise; and optimizing acquisition parameters like pulse sequences and timings for given signals of interest.

The most common form of post-hoc synthetic data validation among the curated literature samples (37 of 39 publications) and self-reported synthetic data applications (9 of 14 publications) was visual inspection. Quantitative validation approaches, including but not limited to comparing specific signal intensities and direct cross-correlation between synthetic and in vivo acquisitions, were in the minority.

## 6.2. Reporting Standards

***Considerations***

- *Synthetic MRS datasets should be stored in a standardized format, such as NIfTI-MRS, to facilitate interoperability, reproducibility, and downstream analysis. A corresponding JSON sidecar file should be provided to document all relevant generation parameters, modeling assumptions, and simulation settings.*
- *A checklist summarizing the minimum recommended considerations, as defined by the respective subgroups, should accompany the dataset or publication to ensure transparency and completeness of reporting.*

Experts' consensus recommendations in the MRS community have established minimum reporting standards for *in vivo* MRS data acquired (MRSinMRS)*. It should be no different for synthetic MRS datasets, where we need to establish reporting and data formatting standards. Synthetic datasets tailored for different studies can vary based on multiple factors, including simulated basis sets (see Section 2.1), optimal signal model settings chosen (see Section 2.2) and metabolite and parameter ranges applied (see Section 2.3).

Depending on the purpose of their study, researchers can either make use of downloadable, ready-made synthetic datasets or generate synthetic datasets using available software. Using downloadable datasets can enable straightforward comparison across literature, whereas generating new synthetic datasets using available software is necessary to investigate new topics. On the other hand, another key factor influencing reproducibility and user-friendliness is the data format of the existing datasets or possible export format options provided in the synthetic dataset generator software. To make informed decisions about a given synthetic dataset or generation software, it is essential to carefully document the data provenance with the corresponding parameters and to report them when presenting or making use of synthetic datasets.

A thorough literature search identified a list of previously used synthetic MRS datasets to investigate current reporting practices. Included within this list are downloadable, ready-made datasets, datasets that were not made publicly available, and software designed to create synthetic datasets. Table S1 Master Table provides a detailed overview of whether parameters used in generating the used synthetic MRS datasets are reported or not in the respective literature. It also indicates the respective data format of the synthetic datasets.

The identified synthetic datasets were produced to be representative of *in vivo* human and rodent MRS data (Gudmundson et al., 2023; Rizzo et al., 2023; Shamaei et al., 2023). These datasets have been used primarily for validating linear combination modeling (Marjańska et al., 2022) and developing machine learning and deep learning algorithms (Gudmundson et al., 2023; Rizzo et al., 2023; Shamaei et al., 2023).

## 6.3. Data Format

Taking the first step towards establishing reporting standards and standardizing a data format for synthetic MRS datasets, in this section, we review the current practices in reporting and data formats that can be generated with the available software.

As an initial step toward establishing reporting standards and standardizing data formats for synthetic MRS datasets, this section reviews the current practices in reporting along with the data formats supported by currently available software. Furthermore, we recommend standardizing data format, and we provide minimum reporting standards for future work presenting synthetic datasets.

Readymade datasets have been created in various formats, including Python NumPy Archive (.npz) files (Gudmundson et al., 2023), Matlab (.dat) (Rizzo et al., 2023), LCModel (.RAW and .h2o) (Marjańska et al., 2022), jMRUI (.txt) (Marjańska et al., 2022), and standard text files (.txt) (Marjańska et al., 2022). These datasets are hosted on Dryad (Gudmundson et al., 2023), MRSHub (Rizzo et al., 2023), and the ISMRM website (Marjańska et al., 2022). Several datasets have also been described within publications for demonstration but not intended to serve as a publicly available dataset. For all of these datasets, nearly all report basic parameters, such as TE/TR, field strength, and the metabolites that were included, as can be seen from the Table S1 Master Table. Many have also detailed out a more extensive list of parameter values.

## 6.4. Challenges and next steps

### 6.4.1. Synthetic Data Validation Methods

Objective, quantitative validation of synthetic MRS data is seldom reported in the literature, making it difficult to gauge the suitability of the generated data in numerous instances. To address this, it is key to first establish basic criteria to validate synthetic data. This may be achieved by evaluating fundamental signal properties and statistics, such as correlations or similarity coefficients to *in vivo* or phantom data, as well as by quantitative outcomes like the *in vivo* performance of models trained on synthetic data. Any such criteria might best be presented as broad guidelines, with sufficient flexibility to accommodate diverse and unforeseen applications. For example, suggesting that synthetic data be validated “by at least one quantitative comparison with similar *in vivo* data” may be a more functional guideline than “frequency-domain correlation of magnitude synthetic spectra with similar *in vivo* data”. The former leaves open the potential for applications centered on time-domain FIDs, real-only spectral analysis, or synthetic data for which one-to-one signal correlations are not the most salient aspect of their utility. Similarly, corresponding guidelines for the systematic reporting of validation techniques, ideally integrated with the broader reporting of synthetic data generation (Section 6.3), would also be beneficial for clarifying the limits of a given synthetic dataset’s generalizability.

### 6.4.2. Reporting Standards

Publications that are focused on the generation or use of synthetic data for other applications in their studies tend to report some details and parameters of synthetic data, but not all that would be required to successfully reproduce the methods from these publications. Overall, parameters that are considered basic, such as field strength, TE, sequence, etc., tend to get reported more consistently across publications. However, more 'sophisticated' parameters related to scaling of the metabolite concentration ranges, adding noise, macromolecules, spline baselines, or artifacts seem very limited. Although these parameters tend to get overlooked, these are equally crucial to report since there are varied methods preferred by different studies to sample and add these other signal components.

With various methods for synthetic MRS data generation, it is critical to define a reporting standard and minimum documentation requirements, similar to the minimum reporting standards for *in vivo* MRS (Lin et al., 2021). As a rule of thumb, all information regarding the simulated basis set, applied signal model parameters, metabolite ranges utilized, and other applied sophisticated parameters must be included. This is imperative as the dataset may be used for validating modeling software, informing supervised neural network training, and identifying areas for improvement. Furthermore, this documentation is needed to understand which dataset is best suited for a given task and to provide source documentation within publications where an extensive list of experimental details cannot be listed. As a guideline, we have created minimum reporting standards for synthetic MRS datasets (MRSsynMRS table as shown in Figure 7 and provided in Supplementary Table S4). This table is intended to be used as a template and can be modified to include/exclude experiment specific details or different metabolites and nuclei as needed.

| **Synthetic MRS Data Reporting Standard** *(Abbreviated)* | | | |
|---|---|---|---|
| **0. Experiment** | | | |
| a. ID | | 1 | |
| **1. Software** | | | |
| a. Basis Set Simulation | | | |
| i. Name | | simulation_name | |
| ii. Repository (Link, DOI, or RRID) | | github.com/simulation | |
| b. Signal Model | | | |
| **. . .** | | | |
| **2. Pulse Sequence** | | | |
| a. Field Strength | [ T ] | 3.0 | |
| b. Localization | | PRESS | |
| **. . .** | | | |
| **3. Metabolites** | | | |
| a. Population | | | |
| i. Age Range | [ yrs ] | 18.1 | 32.6 |
| ii. Healthy Included | | Yes | |
| iii. Clinical Included | | Yes | |
| **. . .** | | | |
| e. Concentration Units | | mM | |
| f. Concentration Ranges | | | |
| - N-acetylaspartate (NAA) | [ mM ] | 9.75 | 18.43 |
| **. . .** | | | |
| **4. Components** | | | |
| **. . .** | | | |
| e. Macromolecule | | | |
| i. Simulated, Modeled, Experimental, or Mixed | | Simulated | |
| ii. Available in Isolation | | Yes | |
| iii. Available as Combination | | Yes | |
| **. . .** | | | |
| **5. Quality** | | | |
| a. Signal to Noise Ratio Range | | 5.72 | 90.39 |
| b. Linewidths | | | |
| i. tCr Linewidth | [ Hz ] | 6.85 | 17.22 |
| ii. tNAA Linewidth | [ Hz ] | 6.91 | 18.07 |
| **. . .** | | | |

**Figure 7.** MRSsynMRS table. An abbreviated version of the Synthetic MRS Data Reporting Standards spreadsheet illustrating: overall structure including 5 major categories; organization layout and formatting to facilitate readability; demonstration of the ability to add values and ranges for a given line item. The full table is available as an excel spreadsheet and includes excel 'notes' in value tabs providing guidance on inputting your own values. While the full table is meant to be inclusive to a variety of niches (e.g., MRSI, dMRS, fMRS, etc.) with approximately 100 inputs and modality-specific ($B_0$ Map, b-values, concentration percent changes, etc.) sections, users are encouraged to add rows as needed to ensure their methods are comprehensively provided for complete transparency and reproducibility. A downloadable version is available here: SyntheticMRS_Reporting_Standard.xlsx

### 6.4.3. Data Format

To create a library of downloadable synthetic datasets for the community, agreeing upon a standard data format as a community is necessary. In addition, this will ensure readability of the data using various modeling software, allowing us to test reproducibility or to compare performance. We recommend exporting and storing the synthetic MRS data in the NifTi-MRS (Clarke et al., 2022) format. To ensure full reproducibility, synthetic datasets should be accompanied by structured metadata stored in a JSON sidecar file alongside the NIfTI data. Field names and nomenclature should follow established standards where available, adopt existing community defined metadata conventions, and remain aligned with the MRSsynMRS reporting structure wherever applicable. The metadata should be sufficiently detailed to satisfy current reporting standards, enabling complete reconstruction of the data generation process.

A major goal for the future from this enormous effort is the creation of standardized tools for generating synthetic data that can be tailored to different research needs, including clinical, preclinical, and methodological studies. Such a tool could help ensure that synthetic data is used more effectively and consistently across the MRS community, ultimately improving the reproducibility and reliability of MRS research.

# 7. Discussion

## 7.1 Standardization and Accessibility

The increasing use of synthetic data in MRS highlights a growing need for shared datasets, common terminology, and transparent simulation practices. Current synthetic data generators differ substantially in how they model basis sets, nuisance signals, relaxation effects, spatial variability, and temporal dynamics. These differences complicate direct comparison of results across studies and limit reproducibility, particularly when synthetic datasets are used for software validation, benchmarking, or algorithm development. As with reporting standards established for experimental spectroscopy, community-driven guidance is needed to harmonize assumptions, parameter definitions, and output formats across simulation frameworks. Standardization efforts analogous to existing reporting initiatives would facilitate comparability while preserving the flexibility required to address diverse experimental questions.

Accessibility is equally important. Synthetic datasets that are publicly available, well documented, and accompanied by clear descriptions of underlying assumptions provide an essential resource for method development, education, and evaluation. Such resources lower barriers to entry, enable fair comparison of processing pipelines, and support reproducible research across institutions and application domains.

## 7.2 Incorporation of Realistic Variability

A central challenge in synthetic spectroscopy is bridging the gap between idealized signal models and the complexity of experimental data. Variability and sparsity of spectral features remain areas of active development. Macromolecular, lipid, and baseline signal generation varies widely across simulators, despite the well-documented influence of broad signal modeling on linear combination modeling outcomes. This raises the possibility that certain simulation strategies may preferentially favor specific quantification algorithms, an issue that warrants systematic investigation.

Beyond broad signals, the mathematical descriptions of eddy currents, spurious echoes, residual water, motion-related artifacts, and other nuisance contributions differ considerably between implementations. Even fundamental metrics such as signal-to-noise ratio are not defined consistently across generators, reflecting both historical conventions and differences in modeling complexity. Some simulators rely on weighted combinations of idealized functions, while others aim to reproduce the underlying physics and chemistry of pulse sequences with detailed timing and spatial information. This diversity reflects the inherent complexity of magnetic resonance experiments but also underscores the need for clearer characterization of model assumptions.

The inclusion of an unsuppressed water reference represents an important step toward improving realism and compatibility with *in vivo* quantification workflows. Generating paired water and metabolite spectra would address ongoing challenges in translating metabolite specific and water-specific relaxation and diffusion properties into appropriate amplitude scaling. Such an approach would also allow direct inversion of water scaling procedures commonly applied in tissue-specific quantification, improving consistency between synthetic and experimental analyses.

Spatial variability remains another limitation. Region-specific metabolite concentrations and relaxation properties are still incompletely characterized, restricting their incorporation into simulation frameworks. Similarly, the integration of $B_0$ and $B_1$ inhomogeneities and coil sensitivity profiles would substantially improve the realism of spatially resolved synthetic data, particularly for spectroscopic imaging applications. Addressing these factors is especially relevant for studies that aim to evaluate spatial quantification accuracy or reconstruction performance.

Finally, there is a growing need for multi-spectrum synthetic datasets that capture temporal structure across acquisitions. Functional spectroscopy, diffusion-weighted spectroscopy, relaxometry, and fingerprinting approaches all require large numbers of temporally related spectra with controlled ground truth. Continued advances in computational resources, algorithmic efficiency, and prior knowledge of acquisition variability will enable more realistic multi-transient simulations that better reflect experimental conditions.

## 7.3 Enhancing Interdisciplinary Collaboration

The development of realistic and reliable synthetic spectroscopy data inherently spans multiple disciplines. Progress depends on coordinated contributions from magnetic resonance physics, neuroscience, clinical research, artificial intelligence, and software engineering. High fidelity synthetic data are particularly critical for artificial intelligence applications, where domain shift between simulated and experimental data can severely limit performance. Carefully designed benchmarks, uncertainty estimates, and standardized reporting are therefore essential components of future simulation frameworks.

Open-source tools, transparent documentation, and shared benchmark datasets provide a foundation for collaboration and collective progress. By aligning simulation strategies with modality-specific requirements and application contexts discussed throughout this work, synthetic data can support a broad range of clinical, preclinical, and methodological studies. Sustained interdisciplinary engagement will be essential to ensure that synthetic spectroscopy continues to evolve in parallel with experimental practice and emerging analytical approaches.

# 8. Conclusion

Synthetic MRS data have become an essential resource for advancing acquisition strategies, processing methods, quantification techniques, and artificial intelligence-based applications. By outlining core, advanced, modality-specific, and application-driven components of simulation frameworks, this work highlights both the substantial progress achieved and the challenges that remain. Although current simulators capture many fundamental aspects of the spectroscopy signal, important sources of variability related to physiology, spatial heterogeneity, temporal dynamics, and hardware influences are still only partially represented.

Continued refinement of simulation models, improved characterization of biological and technical factors, and the integration of unsuppressed water references, voxel-specific field information, and dynamic processes will further enhance realism and applicability. Community-driven efforts toward standardization, transparency, and the sharing of benchmark synthetic datasets are critical for reproducibility and comparability across studies. As synthetic data play an increasingly central role in clinical and preclinical research and in artificial intelligence development, unified and well-validated simulation frameworks will become even more significant. Advancing synthetic spectroscopy therefore requires sustained interdisciplinary collaboration and long-term investment in open, well-documented resources to support robust, accessible, and impactful research across the field.

## Author CRediT Contributions:

**John T. LaMaster**: Conceptualization, Methodology, Investigation, Resources, Data Curation, Writing - Original Draft Preparation, Writing - Review & Editing, Visualization, Supervision, Project administration; **Aaron T. Gudmundson**: Conceptualization, Methodology, Investigation, Resources, Data Curation, Writing - Original Draft Preparation, Writing - Review & Editing, Visualization, Supervision, Project administration; **Alireza Abaei**: Investigation, Writing - Review & Editing; **Seyma Alcicek**: Investigation, Writing - Review & Editing; **Arturo Alvarado** : Investigation, Writing - Review & Editing; **Ovidiu Andronesi**: Conceptualization, Investigation, Writing - Review & Editing; **Tiffany Bell**: Investigation, Data Curation, Writing - Original Draft Preparation, Writing - Review & Editing, Visualization, Project Administration; **Wolfgang Bogner**: Investigation, Writing - Review & Editing, Writing – Original Draft Preparation, Supervision; **Hanna Bugler**: Investigation, Data Curation, Writing - Review & Editing; **Alexander R. Craven**: Investigation, Data Curation, Writing - Original Draft Preparation, Writing - Review & Editing, Project Administration; **Cristina Cudalbu**: Investigation, Writing - Review & Editing; **Alma Davidson**: Investigation, Writing - Review & Editing; **Christopher W. Davies-Jenkins**: Investigation, Writing - Original Draft, Writing - Review & Editing, Visualization; **Dinesh Deelchand**: Investigation, Writing - Review & Editing; **Richard Edden**: Investigation, Writing - Review & Editing; **Morteza Esmaeili**: Investigation, Writing - Review & Editing; **Candace C. Fleischer**: Conceptualization, Investigation, Writing - Original Draft Preparation, Writing - Review & Editing, Supervision; **Abdelrahman Gad**: Investigation, Writing - Review & Editing; **Guglielmo Genovese**: Investigation, Writing - Review & Editing; **Saumya Gurbani**: Resources, Writing - Review & Editing; **Ashley D. Harris**: Investigation, Writing - Review & Editing; **Pierre-Gilles Henry**: Investigation, Writing - Review & Editing; **Kay Chioma Igwe**: Investigation, Data Curation, Writing - Original Draft Preparation, Writing - Review & Editing, Visualization, Project Administration; **Ajin Joy**: Investigation, Writing - Review & Editing; **Margarida Julià-Sapé**: Investigation, Writing – Review & Editing; **Hyeonjin Kim**: Investigation, Writing – Review & Editing; **Roland Kreis**: Investigation, Investigation, Writing - Review & Editing; **Fan Lam**: Investigation, Data Curation, Writing - Review & Editing; **Karl Landheer**: Investigation, Writing - Review & Editing; **Bernard Lanz**: Investigation, Writing - Review & Editing; **Chu-Yu Lee**: Investigation, Writing - Review & Editing; **Clémence Ligneul**: Investigation, Writing - Review & Editing; **Julian P. Merkofer**: Investigation, Data Curation, Writing - Original Draft Preparation, Writing - Review & Editing, Visualization, Project Administration; **Jack J. Miller**: Investigation, Data Curation, Writing - Original Draft Preparation, Writing - Review & Editing, Visualization, Project Administration; **Jessie Mosso**: Methodology, Investigation, Data Curation, Writing - Original Draft Preparation, Writing - Review & Editing, Visualization, Supervision, Project Administration; **Stanislav Motyka**: Investigation, Writing - Review & Editing; **Eloïse Mougel**: Investigation, Writing - Review & Editing; **Paul G. Mullins**: Investigation, Data Curation, Writing - Original Draft, Writing - Review & Editing; **Saipavitra Murali-Manohar**: Investigation, Data Curation, Writing - Original Draft Preparation, Writing - Review & Editing, Project Administration; **Chloé Najac**: Investigation, Writing - Review & Editing; **Shinichiro Nakajima**: Investigation, Writing - Review & Editing; **Georg Oeltzschner**: Conceptualization, Methodology, Investigation, Resources, Writing - Original Draft Preparation, Writing - Review & Editing, Visualization, Project Administration, Supervision; **İpek Özdemir**:

Investigation, Writing - Original Draft Preparation, Writing - Review & Editing; **Esin Ozturk-Isik**: Investigation, Writing - Review & Editing; **Marco Palombo**: Investigation, Writing - Review & Editing; **Ulrich Pilatus**: Investigation, Writing - Review & Editing; **Justyna Platek**: Investigation, Writing - Original Draft Preparation, Writing - Review & Editing; **Esau Poblador Rodriguez**: Investigation, Writing - Review & Editing; **Xiaobo Qu**: Investigation, Writing - Review & Editing; **Rudy Rizzo**: Investigation, Writing - Original Draft Preparation, Writing - Review & Editing; **Christopher T. Rodgers**: Conceptualization, Investigation, Writing - Review & Editing; **Yeison Rodriguez**: Investigation, Writing - Review & Editing; **Manoj K. Sammi**: Investigation, Writing - Review & Editing; **Manoj Kumar Sarma**: Investigation, Data Curation, Writing - Review & Editing; **Francesca Saviola**: Investigation, Writing - Review & Editing; **Anouk Schrantee**: Investigation, Writing - Review & Editing; **Amirmohammad Shamaei**: Data Curation, Investigation, Writing - Review & Editing; **Dunja Simicic**: Investigation, Writing - Review & Editing; **Brian J. Soher**: Conceptualization, Investigation, Writing - Review & Editing; **Nico Sollmann**: Investigation, Writing - Original Draft Preparation, Writing - Review & Editing; **Yulu Song**: Investigation, Writing - Review & Editing; **Jeffrey A. Stanley**: Investigation, Writing - Review & Editing; **Bernhard Strasser**: Investigation, Writing - Original Draft Preparation, Writing - Review & Editing; **Antonia Susnjar**: Investigation, Data Curation, Writing - Original Draft; Writing - Review & Editing, Project Administration; **Kelley M. Swanberg**: Conceptualization, Methodology, Investigation, Resources, Data Curation, Writing - Original Draft Preparation, Writing - Review & Editing, Project administration; **M. Albert Thomas**: Investigation, Writing - Review & Editing; **Ivan Tkáč**: Conceptualization, Investigation, Resources, Data Curation, Writing - Review & Editing; **Zhangren Tu**: Investigation, Writing - Original Draft Preparation, Writing - Review & Editing; **Dennis M. J. van de Sande**: Investigation, Data Curation, Writing - Original Draft Preparation, Writing - Review & Editing, Visualization, Project Administration; **Emma Van Praagh**: Data Curation, Investigation, Writing - Review & Editing; **Paul J. Weiser**: Investigation, Writing - Review & Editing; **Mark Widmaier**: Investigation, Data Curation, Writing - Original Draft Preparation, Writing - Review & Editing, Visualization, Project Administration; **Martin Wilson**: Investigation, Writing - Original Draft Preparation, Writing - Review & Editing, Visualisation, Project Administration; **Christopher J. Wu**: Investigation, Writing - Review & Editing; **Lijing Xin**: Investigation, Writing - Review & Editing; **Helge J. Zöllner**: Investigation, Writing - Original Draft Preparation, Writing - Review & Editing; **Antonia Kaiser**: Conceptualization, Methodology, Investigation, Resources, Data Curation, Writing - Original Draft Preparation, Writing - Review & Editing, Visualization, Supervision, Project administration.

## Data Availability Statement:

The data that supports the findings of this study are available in the supplementary material of this article or are already publicly available.

## Conflicts of Interest:

The authors declare no conflicts of interest.

## Funding:

This work was supported in part by the following grants from the National Institutes of Health: K99 AG080084, K99 EB034768, P30 NS076408, P41 EB027061, P41 EB031771, R00 AG02230, R01 CA211080, R01 CA255479, R01 EB016089, R01 EB023963, R01 EB031787, R01 EB032788, R01 EB035529, and R21 EB033516. This work was supported in part by the following grant from the Schweizerischer Nationalfonds zur Förderung der Wissenschaftlichen Forschung: 225289. This work was supported in part by the following grant from the UK Research and Innovation Future Leaders Fellowships: MR/T020296/2 and UKRI1073. This work was supported in part by the following grant from the UK Research and Innovation Medical Research Council Research: MR/W031566/1. This work was supported in part by the following grants from the Japan Society for the Promotion of Science: 22H03002, 23K24263, 24K02541, 24K02387, 24K10740, 25K21813, 25K10856, and 25K10823. This work was supported in part by the following grants from the Japan Agency for Medical Research and Development: JP24wm0625302 and JP24wm0625307. This work was supported in part by the Christian Doppler Society (BIOMAK). This work was supported in part by the following grant from the HORIZON EUROPE European Research Council: 101088351. This work was supported in part by the following grants from the Austrian Science Fund: I6037, P34198, P36328, and PAT4912324.